\documentclass[%
 aip,
rsi,%
 amsmath,amssymb,
reprint,%
floatfix,
]{revtex4-1}

\usepackage{graphicx}
\usepackage{dcolumn}
\usepackage{braket}
\usepackage{bm}
\usepackage{multirow}
\usepackage{csquotes}
\usepackage{algpseudocode}
\usepackage{amsmath}

\newcommand{\indnuci}[0]{I}
\newcommand{\indnucii}[0]{J}
\newcommand{\indocci}[0]{i}
\newcommand{\indoccii}[0]{j}

\newcommand{\indx}[0]{a}
\newcommand{\indy}[0]{b}

\newcommand{\then}[1]{N_\mathrm{#1}}
\newcommand{\thei}[0]{\mathrm{i}}
\newcommand{\theone}[0]{\mathbf{1}}
\newcommand{\theelcharge}[0]{e}

\newcommand{\ther}[1]{\mathbf{R}_{#1}}
\newcommand{\therel}[2]{R_{#1}^{#2}}
\newcommand{\therbar}[2]{\mathbf{\bar{R}}_{#1#2}}
\newcommand{\them}[1]{\ifthenelse{\equal{#1}{}}{\mathbf{M}}{M_{#1}}}
\newcommand{\therdot}[1]{\dot{\mathbf{R}}_{#1}}
\newcommand{\therdotdir}[2]{\dot{R}_{#1}^\mathrm{#2}}
\newcommand{\therddot}[1]{\ddot{\mathbf{R}}_{#1}}
\newcommand{\thelen}[2]{d_{#1#2}}
\newcommand{\theorig}[0]{\mathbf{O}}

\newcommand{\thez}[1]{Z_{#1}}
\newcommand{\theb}[1]{\ifthenelse{\equal{#1}{}}{\mathbf{B}}{B_\mathrm{#1}}}
\newcommand{\thebt}[1]{\ifthenelse{\equal{#1}{}}{\mathbf{\tilde{B}}}{\tilde{B}_{#1}}}

\newcommand{\thef}[2]{\mathbf{F}_{#1}^\mathrm{#2}}
\newcommand{\thefdir}[3]{F_{#1}^\mathrm{#2#3}}

\newcommand{\theel}[1]{\mathbf{r}_{#1}}

\newcommand{\themo}[2]{\ifthenelse{\equal{#2}{1}}{\varphi_{#1} (\theel{};\ther{},\theorig{},\theb{})}{\varphi_{#1}}}

\newcommand{\themomul}[3]{\ifthenelse{\equal{#3}{1}}{\varphi_{#1}^{(#2)} (\theel{};\ther{},\theorig{})}{\varphi_{#1}^{(#2)}}}

\newcommand{\thefluc}[4]{\ifthenelse{\equal{#4}{0}}
{\boldsymbol{\alpha}_{#1#2} (#3)}
{[\boldsymbol{\alpha}_{#1#2} (#3)]^\mathrm{T}}
}

\newcommand{\theapt}[3]{\ifthenelse{\equal{#3}{0}}
{\mathbf{V}_{#1} (#2)}
{[\mathbf{V}_{#1} (#2)]^\mathrm{T}}
}

\newcommand{\theflucs}[4]{\ifthenelse{\equal{#4}{0}}
{\boldsymbol{\alpha}_{#1#2}^{+} (#3)}
{[\boldsymbol{\alpha}_{#1#2}^{+} (#3)]^\mathrm{T}}
}

\newcommand{\thefluca}[4]{\ifthenelse{\equal{#4}{0}}
{\boldsymbol{\alpha}_{#1#2}^{-} (#3)}
{[\boldsymbol{\alpha}_{#1#2}^{-} (#3)]^\mathrm{T}}
}

\newcommand{\thee}[2]{#1_\mathrm{#2}}

\newcommand{\theom}[3]{\boldsymbol{\Omega}_{#1#2} (#3)}
\newcommand{\theomel}[5]{\Omega_{#1#2}^{#3#4} (#5)}

\newcommand{\thechi}[2]{\boldsymbol{\chi}_{#1} (#2)}
\newcommand{\thechiel}[3]{\chi_{#1}^{#2} (#3)}

\newcommand{\theoms}[3]{\boldsymbol{\Omega}_{#1#2}^{-} (#3)}
\newcommand{\theoma}[3]{\boldsymbol{\Omega}_{#1#2}^{+} (#3)}

\newcommand{\theomsym}[4]{\boldsymbol{\Omega}_{#1#2}^\mathrm{#4} (#3)}
\newcommand{\thelamsym}[4]{\boldsymbol{\Lambda}_{#1#2}^\mathrm{#4} (#3)}

\newcommand{\thecharge}[3]{q_{#1}^\mathrm{#2} (#3)}

\newcommand{\theeta}[2]{\eta_{#1#2} (\ther{})}

\newcommand{\theq}[3]{Q_{#1#2} (#3)}
\newcommand{\thep}[3]{P_{#1#2} (#3)}

\newcommand{\theqm}[3]{Q_{#1#2}^\mathrm{M} (#3)}

\newcommand{\themu}[2]{\boldsymbol{\mu}_{#1} (#2)}

\newcommand{\thepelop}[1]{\mathbf{\hat{p}}_{#1}}
\newcommand{\thepnucop}[1]{\mathbf{\hat{P}}_{#1}}
\newcommand{\thepnucdir}[2]{\hat{P}_{#1}^{#2}}

\newcommand{\thegas}[3]{\boldsymbol{\Gamma}_{#1#2}^+ (#3)}

\begin{document}

\title[]{Berry Population Analysis:\\ Atomic Charges from the Berry Curvature in a Magnetic Field}

\author{Laurens D. M. Peters}
\email{laurens.peters@kjemi.uio.no}
\affiliation
{Hylleraas Centre for Quantum Molecular Sciences,  Department of Chemistry, 
University of Oslo, P.O. Box 1033 Blindern, N-0315 Oslo, Norway}
\author{Tanner Culpitt}
\affiliation
{Hylleraas Centre for Quantum Molecular Sciences,  Department of Chemistry, 
University of Oslo, P.O. Box 1033 Blindern, N-0315 Oslo, Norway}
\author{Erik I. Tellgren}
\affiliation
{Hylleraas Centre for Quantum Molecular Sciences,  Department of Chemistry, 
University of Oslo, P.O. Box 1033 Blindern, N-0315 Oslo, Norway}
\author{Trygve Helgaker}
\affiliation
{Hylleraas Centre for Quantum Molecular Sciences,  Department of Chemistry, 
University of Oslo, P.O. Box 1033 Blindern, N-0315 Oslo, Norway}

\date{\today}

\begin{abstract}
The Berry curvature is essential in Born--Oppenheimer molecular dynamics, describing the screening of the nuclei by the electrons in a magnetic field. Parts of the Berry curvature can be understood as the external magnetic field multiplied by an effective charge so that the resulting Berry force behaves like a Lorentz force during the simulations. Here we investigate whether these effective charges can provide insight into the electronic structure of a given molecule or, in other words, whether we can perform a population analysis based on the Berry curvature. To develop our approach, we first rewrite the Berry curvature in terms of charges that partially capture the effective charges and their dependence on the nuclear velocities. With these Berry charges and charge fluctuations, we then construct our population analysis yielding atomic charges and overlap populations. Calculations at the Hartree--Fock level reveal that the atomic charges are similar to those obtained from atomic polar tensors. However, since we additionally obtain an estimate for the fluctuations of the charges and a partitioning of the atomic charges into contributions from all atoms, we conclude that the Berry population analysis is a useful alternative tool to analyze the electronic structure of molecules.\\
\end{abstract}

\maketitle

\section{Introduction}

Population analysis is one of the simplest and most common tools of a quantum chemist to gain insight into the electronic structure of a molecular system.\cite{Wiberg1993,Meister1994,Cramer2004,Cho2020} The central idea is that we can assign a partial or atomic charge $q_\indnuci$ to every atom $\indnuci$, allowing us to analyze the bonding situation or to make predictions regarding the reactivity of a compound, avoiding a more complex analysis of the electronic density.

Today, a large number of population analyses exist, each with its own pros and cons. Many methods rely on a partitioning of either the wave function or the electron density into atomic fragments. Prominent examples are the Mulliken,\cite{Mulliken1955,Mulliken1955a,Mulliken1955b} L\"owdin,\cite{Lowdin1950,Baker1985} natural,\cite{Reed1985} Bader,\cite{Maslen1985} and Hirshfeld\cite{Hirshfeld1977} charges. While the first two use atom-centered basis functions to determine an atomic contribution to the wave function, the remaining directly determine atomic densities that add up to the total electronic density. At this point, we should mention the Charge Model 5 (CM5\cite{Marenich2012}) and the Density Derived Electrostatic and Chemical (DDEC6\cite{Manz2016,Limas2016}) charges as significant improvements on these methods.

The second large group of population analyses tries to access the atomic charges via observables of the molecule. Restrained Electrostatic Potential (RESP) charges\cite{Bayly1993} are, for example, extracted from the electrostatic potential of the molecule, whereas the atomic-polar-tensor or Born effective charges are determined as derivatives of the electronic dipole moment with respect to nuclear displacements.\cite{Cioslowski1989,Cioslowski1990,Haaland2000,Shukla2000,Milani2010} 

Here, we introduce a new population analysis that fits into the second group discussed above. The idea behind our method is simple: Any charged particle moving in a magnetic field $\theb{}$ with velocity $\therdot{\indnuci}$ experiences a Lorentz force $\mathbf{F}_\indnuci^\mathrm{L}$ inducing a cyclic motion about the magnetic field vector,
\begin{equation}
\mathbf{F}_\indnuci^\mathrm{L} = q_\indnuci \theb{} \times \therdot{\indnuci} .
\label{int_000}
\end{equation}
Consequently, when the forces acting on the nuclei of a molecule in a magnetic field are known, we can use these to determine atomic charges.

The \textit{ab initio} calculation of forces of molecules in a magnetic field is, however, not straightforward; only very recently\cite{Peters2021,Monzel2022} were the first simulations conducted. They require a non-perturbative treatment of the magnetic field\cite{Tellgren2008,Tellgren2009,Lange2012,Tellgren2012,Reynolds2015,Stopkowicz2015,Hampe2017,Irons2017,Hampe2019,Sen2019,Sun2019,Austad2020,Hampe2020,Pausch2020,Williams-Young2020,Irons2021,Blaschke2022} and the use of London orbitals\cite{London1937,Hameka1958,Ditchfield1976,Helgaker1991} to ensure the correct physics -- namely, that all observables are gauge and translationally invariant\cite{Culpitt2021} and that neutral atoms and molecules do not \enquote{feel} an overall Lorentz force in a magnetic field.\cite{Peters2022}  In a magnetic field, each nucleus experiences not only the usual Lorentz force, but also the Berry force, generated by the electrons in the system\cite{Schmelcher1988,Schmelcher1989,Yin1992,Peternelj1993,Yin1994,Schmelcher1997,Ceresoli2007}
\begin{equation}
\mathbf{F}_\indnuci^\mathrm{B} =
\sum \limits_{\indnucii=1}^{\then{nuc}} \boldsymbol{\Omega}_{\indnuci\indnucii}  \therdot{\indnucii} ,
\label{int_001}
\end{equation}
where $\boldsymbol{\Omega}_{\indnuci\indnucii}$ is the Berry curvature,\cite{Berry1984,Mead1992,Anandan1997,Resta2000} representing the screening of the nuclei by the electrons. It contains derivatives of the electronic wave function with respect to the nuclear coordinates and can be determined at the Hartree--Fock (HF) level of theory, via a numerical scheme\cite{Culpitt2021} or by solving the coupled-perturbed HF equations.\cite{Culpitt2022}

In this work, we demonstrate how $\boldsymbol{\Omega}_{\indnuci\indnucii}$ can be used as a population analysis of a given molecule. After a short discussion of its calculation and properties (Section~II.A), we rewrite the Berry curvature in terms of polarization tensors (Section~II.B) and Berry charges (Section~II.C). The latter give a simple picture of the screening process captured by the Berry curvature (Section~II.D), which we use as a justification for the Berry population analysis (Section~II.E). Having summarized the computational details in Section~III, we analyze the Berry charges and validate the resulting  population analysis via comparison to the established Mulliken\cite{Mulliken1955} and atomic-polar-tensor\cite{Cioslowski1989} charges in Section~IV. Conclusions and future directives are given in Section~V. 

\section{Theory}

We use indices $\indx, \indy, ...$ for Cartesian components, indices $\indnuci, \indnucii, ...$ for the $\then{nuc}$ nuclei, and indices $\indocci, \indoccii, ...$ for the $\then{occ}$ occupied molecular orbitals $\{\themo{\indocci}{1}\}$. The electronic and nuclear coordinates are denoted by $\theel{}$ and $\ther{}$, respectively, while $\thepelop{}$ and $\thepnucop{}$ refer to the corresponding canonical momentum operators:
\begin{align}
\thepelop{}  &= - \thei{} \hbar \dfrac{\partial}{\partial \theel{}}, \qquad
\thepnucop{} = - \thei{} \hbar \dfrac{\partial}{\partial \ther{}}. \label{not_001} 
\end{align}
We use $\theelcharge{}\thez{\indnuci}$, $\them{\indnuci}$, $\ther{\indnuci}$, and $\therdot{\indnuci}$ to represent the charge, mass, coordinates, and velocity of nucleus $\indnuci$. Here $\theorig{}$ is the gauge origin and $\theb{}$ a uniform magnetic field. By introducing the magnetic field tensor
\begin{align}
\thebt{} =
\begin{pmatrix} 
0 & - \theb{z} & \theb{y} \\
\theb{z} & 0 & -\theb{x} \\
-\theb{y} & \theb{x} & 0
\end{pmatrix}, \qquad \theb{} =
\begin{pmatrix} 
\theb{x} \\
\theb{y} \\
\theb{z} 
\end{pmatrix} ,
\label{not_002}
\end{align}
we may reformulate a cross product of $\mathbf B$ with a vector $\mathbf v$ as a matrix multiplication $\mathbf B \times \mathbf v = \thebt{} \mathbf v$.
We also define the Jacobian matrix for derivatives of vectors with respect to a nuclear coordinate by
\begin{align}
\frac{\partial \mathbf{W}_{\indnucii} (\ther{})}{\partial \ther{\indnuci}} =
\begin{pmatrix}
\dfrac{\partial W_{J\mathrm{x}} (\ther{})}{\partial R_{I\mathrm{x}}} & \dfrac{\partial W_{J\mathrm{x}} (\ther{})}{\partial R_{I\mathrm{y}}} & \dfrac{\partial W_{J\mathrm{x}} (\ther{})}{\partial R_{I\mathrm{z}}} \\
\dfrac{\partial W_{J\mathrm{y}} (\ther{})}{\partial R_{I\mathrm{x}}} & \dfrac{\partial W_{J\mathrm{y}} (\ther{})}{\partial R_{I\mathrm{y}}} & \dfrac{\partial W_{J\mathrm{y}} (\ther{})}{\partial R_{I\mathrm{z}}} \\
\dfrac{\partial W_{J\mathrm{z}} (\ther{})}{\partial R_{I\mathrm{x}}} & \dfrac{\partial W_{J\mathrm{z}} (\ther{})}{\partial R_{I\mathrm{y}}} & \dfrac{\partial W_{J\mathrm{z}} (\ther{})}{\partial R_{I\mathrm{z}}}
\end{pmatrix}.
\label{not_003}
\end{align}
For brevity, we drop the dependence of operators, expectation values, and orbitals on $\theorig{}$ and $\theb{}$.

\subsection{Berry Curvature in a Magnetic Field}

The force on atom $I$ in a magnetic field consists of the Born--Oppenheimer force, the Lorentz force, and the Berry force:\cite{Yin1992,Ceresoli2007,Culpitt2021,Peters2021,Culpitt2022}
\begin{align}
\thef{\indnuci}{} =
\them{\indnuci} \therddot{\indnuci}  =
 - \dfrac{\partial \thee{E}{BO}(\ther{})}{\partial \ther{\indnuci}} &-
\thez{\indnuci} \theelcharge{}  \thebt{} \therdot{\indnuci} \nonumber \\&
+ \sum \limits_{\indnucii=1}^{\then{nuc}} \theom{\indnuci}{\indnucii}{\ther{}}  \therdot{\indnucii} .
\label{cur_000}
\end{align}
Here, $\thee{E}{BO}(\ther{})$ is the potential energy with or without the diagonal Born--Oppenheimer correction (DBOC) included and $\theom{\indnuci}{\indnucii}{\ther{}}$ is the Berry curvature, which is determined from derivatives of the geometric vector potential [$\thechi{\indnuci}{\ther{}}$] with respect to the nuclear coordinates:
\begin{align}
\theom{\indnuci}{\indnucii}{\ther{}}
&= 
\dfrac{\partial \thechi{\indnuci}{\ther{}}}{\partial \ther{\indnucii}} - 
\left[
\dfrac{\partial \thechi{\indnucii}{\ther{}}}{\partial \ther{\indnuci}} 
\right]^\mathrm{T}. \label{cur_000b} 
\end{align}
In HF and density-functional-theory (DFT) calculations, both quantities are calculated from derivatives of the occupied molecular orbitals with respect to the nuclear coordinates:
\begin{align}
\thechiel{\indnuci}{\indx}{\ther{}}
&= 
\sum \limits_{\indocci=1}^{\then{occ}}
\Braket{
\themo{\indocci}{0} (\ther{})
|
\thepnucdir{\indnuci}{\indx} \themo{\indocci}{0} (\ther{})
},
\label{cur_001a} \\
\theomel{\indnuci}{\indnucii}{\indx}{\indy}{\ther{}}
&= 
- \dfrac{2}{\hbar} \Im \Bigg \{
\sum \limits_{\indocci=1}^{\then{occ}}
\Braket{
\thepnucdir{\indnuci}{\indx} \themo{\indocci}{0} (\ther{})
|
\thepnucdir{\indnucii}{\indy} \themo{\indocci}{0} (\ther{})
}
\Bigg \}.
\label{cur_001b}
\end{align}
More details on their interpretation and calculation from London orbitals are given elsewhere.\cite{Yin1992,Ceresoli2007,Culpitt2021,Peters2021,Culpitt2022,Peters2022} Here, we note that the geometric vector potential is real-valued and gauge dependent, while the Berry curvature has the units $\big[B_0 \theelcharge{}\big]$ (magnetic field strength times electronic charge) and is linked to the screening of the nuclear charges by the electrons in a magnetic field. The components of the Berry curvature obey the magnetic-translational sum rule\cite{Yin1992,Peters2022}
\begin{align}
\sum \limits_{\indnuci,\indnucii=1}^{\then{nuc}} 
\theom{\indnuci}{\indnucii}{\ther{}} 
&= \theelcharge{} \then{elec} \thebt{} ,
\label{cur_002}
\end{align}
and are, by construction, antisymmetric upon permutation of nuclei:
\begin{align}
\theom{\indnuci}{\indnucii}{\ther{}} &= -
\big[ \theom{\indnucii}{\indnuci}{\ther{}} \big]^\mathrm{T} .
\label{cur_003}
\end{align}
Using the latter property, we can separate each component into parts that are permutationally symmetric ($+$) and antisymmetric ($-$), respectively:
\begin{align}
\theom{\indnuci}{\indnucii}{\ther{}} &= 
\theoms{\indnuci}{\indnucii}{\ther{}} 
+
\theoma{\indnuci}{\indnucii}{\ther{}} ,\label{cur_004} \\
\theoma{\indnuci}{\indnucii}{\ther{}} &= \dfrac{1}{2} \big\{ \theom{\indnuci}{\indnucii}{\ther{}} + \theom{\indnucii}{\indnuci}{\ther{}}  \big\},
\nonumber \\
&= \dfrac{1}{2} \bigg\{
\theom{\indnuci}{\indnucii}{\ther{}} -
\big[ \theom{\indnuci}{\indnucii}{\ther{}} \big]^\mathrm{T}
\bigg\}
\label{cur_005}
\\
\theoms{\indnuci}{\indnucii}{\ther{}} &= \dfrac{1}{2} \big\{ \theom{\indnuci}{\indnucii}{\ther{}} - \theom{\indnucii}{\indnuci}{\ther{}}  \big\}\nonumber \\&= \dfrac{1}{2} \bigg\{
\theom{\indnuci}{\indnucii}{\ther{}} +
\big[ \theom{\indnuci}{\indnucii}{\ther{}} \big]^\mathrm{T}
\bigg\}.
\label{cur_006}
\end{align}

The unit of the Berry curvature and of Eq.~\eqref{cur_002} indicate that (at least parts of) $\theom{\indnuci}{\indnucii}{\ther{}}$ can be interpreted as the external magnetic field multiplied with an effective atomic charge,
\begin{align}
\theom{\indnuci}{\indnucii}{\ther{}} 
&\approx - \thecharge{\indnuci\indnucii}{eff}{\ther{}} \thebt{} ,
\label{cur_007}
\end{align}
where all charges sum up to the total number of electrons:
\begin{align}
\sum \limits_{\indnuci,\indnucii=1}^{\then{nuc}} 
\thecharge{\indnuci\indnucii}{eff}{\ther{}} &= - \theelcharge{}\then{elec}.
\label{cur_008}
\end{align}
Our task is therefore now to rewrite or approximate the Berry curvature in terms of these charges. This will be done in two steps: (1) by separating $\theom{\indnuci}{\indnucii}{\ther{}}$ into a magnetic-field-dependent part and a polarization tensor (Section~II.B); and (2) by reducing the latter to charges (Section~II.C). Having analyzed and visualized these charges (Section~II.D), we construct our population analysis (Section~II.E).

\subsection{Polarization Tensor Approximation of the Berry Curvature}

Let us assume that the geometric vector potential can be approximated in the following manner:
\begin{align}
\thechi{\indnuci}{\ther{}} \approx 
- \dfrac{1}{2} \theb{} \times \themu{\indnuci}{\ther{}} +
\dfrac{\partial \theeta{}{}}{\partial \ther{\indnuci}}.
\label{apt_000}
\end{align}
Here, $\theeta{}{}$ is an arbitrary gauge function and we refer to $\themu{\indnuci}{\ther{}}$ as a \emph{nuclear contribution} to the total electronic dipole moment $\themu{}{\ther{}}$:
\begin{align}
\sum \limits_{\indnuci=1}^{\then{nuc}} \themu{\indnuci}{\ther{}} = \themu{}{\ther{}}.
\label{apt_001}
\end{align}
This relation ensures the correct translational behavior of the total geometric vector potential [$\sum_{\indnuci=1}^{\then{nuc}} \thechi{\indnuci}{\ther{}}$, see ref.~\onlinecite{Peters2022}] and coincides with the dipolar sum rule introduced by Zabalo, Dreyer, and Stengel in Ref.~\onlinecite{Zabalo2022}. Note the analogy between our formulation of $\thechi{\indnuci}{\ther{}}$ and the external vector potential in the Coulomb gauge $[\mathbf{A}_\indnuci (\ther{})]$ as well as the vector potential of a single London orbital centered at $\ther{\indnuci}$ [$\thechi{\mathrm{LDN}}{\ther{\indnuci}}$]\cite{Culpitt2021}:
\begin{align}
\mathbf{A}_\indnuci (\ther{}) &= -\dfrac{\theelcharge{}}{2}
\thez{\indnuci} \theb{} \times (\ther{\indnuci} - \theorig{}),
\label{apt_002a} \\
\thechi{\mathrm{LDN}}{\ther{\indnuci}} &= \dfrac{\theelcharge{}}{2}
\theb{} \times  (\ther{\indnuci} - \theorig{}).
\label{apt_002b}
\end{align}

Inserting our ansatz into Eq.~\eqref{cur_000b} and introducing the polarization tensor,
\begin{align}
\thefluc{\indnuci}{\indnucii}{\ther{}}{0} &= \dfrac{\partial \themu{\indnuci}{\ther{}} }{\partial \ther{\indnucii}} ,
\label{apt_003}
\end{align}
we obtain a compact expression that separates the \emph{explicit} magnetic field dependence ($\thebt{}$, $[B_0]$) from the electronic structure dependence ($\thefluc{\indnuci}{\indnucii}{\ther{}}{0}$, $[\theelcharge{}]$) that depends only \emph{implicitly} on $\theb{}$:
\begin{align}
\theomsym{\indnuci}{\indnucii}{\ther{}}{PT} =  - \dfrac{\theelcharge{}}{2} \big[
\thebt{} \thefluc{\indnuci}{\indnucii}{\ther{}}{0} + \thefluc{\indnucii}{\indnuci}{\ther{}}{1} \thebt{} \big]
\label{apt_009}
\end{align}
From now on, we will refer to $\theomsym{\indnuci}{\indnucii}{\ther{}}{PT}$ as the polarization-tensor approximation of the Berry curvature $\theom{\indnuci}{\indnucii}{\ther{}} \approx 
\theomsym{\indnuci}{\indnucii}{\ther{}}{PT}$.
Note that contributions from the gauge functions vanish, since
\begin{align}
\dfrac{\partial \theeta{}{}}{\partial \therel{\indnuci}{\indx}\partial \therel{\indnucii}{\indy}} - 
\dfrac{\partial \theeta{}{}}{\partial \therel{\indnucii}{\indy}\partial \therel{\indnuci}{\indx}} = 0 .
\label{apt_004}
\end{align}

The approximate Berry curvature $\theomsym{\indnuci}{\indnucii}{\ther{}}{PT}$ retains all the important properties of the exact Berry curvature: (1) the permutational antisymmetry [see eq.~\eqref{cur_003}], (2) the near-linear dependence on the magnetic field strength, and (3) the magnetic-translational sum rule [see Eq.~\eqref{cur_002}]. The latter property can be easily demonstrated by writing out the sum over all polarization tensors:
\begin{align}
\sum \limits_{\indnuci,\indnucii=1}^{\then{nuc}} 
\thefluc{\indnuci}{\indnucii}{\ther{}}{0} &= 
\sum \limits_{\indnucii=1}^{\then{nuc}} \dfrac{\partial \themu{}{\ther{}} }{\partial \ther{\indnucii}} = - \then{elec} \, \theone{}.
\label{apt_011}
\end{align}
This underscores the relation between the dipolar\cite{Zabalo2022} and the magnetic-translational\cite{Peters2022} sum rule. It also shows that $\thefluc{\indnuci}{\indnucii}{\ther{}}{0}$ can be regarded as a \emph{nuclear contribution} to the atomic polar tensor $\theapt{\indnucii}{\ther{}}{0}$:
\begin{align}
\theapt{\indnucii}{\ther{}}{0} &= 
\sum \limits_{\indnuci}^{\then{nuc}} 
\thefluc{\indnuci}{\indnucii}{\ther{}}{0} .
\label{apt_011b}
\end{align}
Note that the polarization-tensor approximation of the Berry curvature in Eq.~\eqref{apt_009} has been used previously in the M3 model in Ref.~\onlinecite{Peters2022}, where $\themu{\indnuci}{\ther{}}$ was obtained from the Mulliken partitioning scheme. Here, we do not assume a \emph{specific} form for $\themu{\indnuci}{\ther{}}$ or $\thefluc{\indnuci}{\indnucii}{\ther{}}{0}$, but use their properties and physical interpretations to gain insight into the Berry curvature itself.

As a final step of the subsection, we write out the permutationally symmetric ($+$) and antisymmetric parts ($-$) of $\theomsym{\indnuci}{\indnucii}{\ther{}}{PT}$,
\begin{align}
\theomsym{\indnuci}{\indnucii}{\ther{}}{PT+} &=
- \dfrac{\theelcharge{}}{2} \big[
\thebt{} \theflucs{\indnuci}{\indnucii}{\ther{}}{0} + \theflucs{\indnuci}{\indnucii}{\ther{}}{1} \thebt{} \big] ,
\label{apt_012}
\\
\theomsym{\indnuci}{\indnucii}{\ther{}}{PT-} &=
- \dfrac{\theelcharge{}}{2} \big[
\thebt{} \thefluca{\indnuci}{\indnucii}{\ther{}}{0} - \thefluca{\indnuci}{\indnucii}{\ther{}}{1} \thebt{} \big] ,
\label{apt_013}
\end{align}
in terms of the corresponding components of the polarization tensor:
\begin{align}
\theflucs{\indnuci}{\indnucii}{\ther{}}{0} = \dfrac{1}{2} \big[ 
\thefluc{\indnuci}{\indnucii}{\ther{}}{0} +
\thefluc{\indnucii}{\indnuci}{\ther{}}{0} 
\big],
\label{apt_014}
\\
\thefluca{\indnuci}{\indnucii}{\ther{}}{0} = \dfrac{1}{2} \big[ 
\thefluc{\indnuci}{\indnucii}{\ther{}}{0} -
\thefluc{\indnucii}{\indnuci}{\ther{}}{0} 
\big].
\label{apt_015}
\end{align}

\subsection{Charge Approximations to the Berry Curvature}

The polarization-tensor approximation to the Berry curvature can be simplified further when we reduce the polarization tensors to a few meaningful components or charges. Here, we focus on two charges: $\theq{\indnuci}{\indnucii}{\ther{}}$ and $\thep{\indnuci}{\indnucii}{\ther{}}$. 

We start by only taking into account the \emph{isotropic} part $[\theq{\indnuci}{\indnucii}{\ther{}}]$ of the permutationally symmetric polarization tensor:
\begin{align}
\theelcharge{} \theflucs{\indnuci}{\indnucii}{\ther{}}{0} &\approx  \dfrac{\theelcharge{}}{3} \mathrm{Tr} \left[ \theflucs{\indnuci}{\indnucii}{\ther{}}{0}\right]\, \theone{} =
\theq{\indnuci}{\indnucii}{\ther{}} \, \theone{} 
\label{cha_000}
\end{align}
This approximation is exact for atoms and dissociated molecules, where the electrons screen the nuclear charges isotropically:
\begin{align}
\theom{\indnuci}{\indnuci}{\ther{}} 
&= - \theelcharge{}\thez{\indnuci}\thebt{} \iff
\theq{\indnuci}{\indnuci}{\ther{}} = - \theelcharge{}\thez{\indnuci} .
\label{cha_001}
\end{align}
It may hold also for highly symmetric molecular systems. From Eqs.~\eqref{apt_011} and \eqref{apt_014}, we know that these charges are permutationally symmetric and sum up to the total number of electrons:
\begin{align}
\theq{\indnuci}{\indnucii}{\ther{}} &= \theq{\indnucii}{\indnuci}{\ther{}}, \quad
\sum \limits_{\indnuci,\indnucii=1}^{\then{nuc}} \theq{\indnuci}{\indnucii}{\ther{}} = - \theelcharge{} \then{elec}.
\label{cha_002}
\end{align}

Additionally, we may reduce the permutationally antisymmetric polarization tensor to its component $\thep{\indnuci}{\indnucii}{\ther{}}$ along the normalized inter-atomic distance vector $\therbar{\indnuci}{\indnucii}$:
\begin{align}
\theelcharge{} \thefluca{\indnuci}{\indnucii}{\ther{}}{0} &\approx  2 \thep{\indnuci}{\indnucii}{\ther{}}
\therbar{\indnuci}{\indnucii}
\therbar{\indnuci}{\indnucii}^\mathrm{T},
\label{cha_003}
\end{align}
where
\begin{align}
 \therbar{\indnuci}{\indnucii} &= 
\dfrac{\ther{\indnucii} - \ther{\indnuci}}{\thelen{\indnuci}{\indnucii}},
\quad
\thelen{\indnuci}{\indnucii} = |\ther{\indnucii} - \ther{\indnuci}|
.
\label{cha_004}
\end{align}
This approximation is exact when the contributions to the electronic dipole moment $\themu{\indnuci}{\ther{}}$ and $\themu{\indnucii}{\ther{}}$  depend solely on $\thelen{\indnuci}{\indnucii}$, while their sum is aligned with $\therbar{\indnuci}{\indnucii}$:
\begin{align}
\theelcharge{} \thefluca{\indnuci}{\indnucii}{\ther{}}{0} &=  
\dfrac{\theelcharge{}}{2} \left[
\dfrac{\partial \themu{\indnuci}{\ther{}} }{\partial \ther{\indnucii}}-
\dfrac{\partial \themu{\indnucii}{\ther{}} }{\partial \ther{\indnuci}}
\right] \nonumber \\
&= \dfrac{\theelcharge{}}{2} \dfrac{\partial }{\partial \thelen{\indnuci}{\indnucii}}  
\left[
\themu{\indnuci}{\ther{}} + \themu{\indnucii}{\ther{}} 
\right] \therbar{\indnuci}{\indnucii}^\mathrm{T}\nonumber \\&= 2 \thep{\indnuci}{\indnucii}{\ther{}}
\therbar{\indnuci}{\indnucii}
\therbar{\indnuci}{\indnucii}^\mathrm{T}.
\label{cha_005}
\end{align}
We therefore expect it to hold for linear molecules with cylindrical symmetry and a non-zero electric dipole moment. Since
$\thefluca{\indnuci}{\indnucii}{\ther{}}{0}$ is permutationally antisymmetric and adds up to zero when summed over all pairs of nuclei, the same relations hold for $\thep{\indnuci}{\indnucii}{\ther{}}$:
\begin{align}
\thep{\indnuci}{\indnucii}{\ther{}} &= - \thep{\indnucii}{\indnuci}{\ther{}}, \quad
\sum \limits_{\indnuci,\indnucii=1}^{\then{nuc}} \thep{\indnuci}{\indnucii}{\ther{}} = 0.
\label{cha_006}
\end{align}

Equipped with Eqs.~\eqref{cha_000} and \eqref{cha_003}, we now construct two charge approximations, termed B1 and B2, to the Berry curvature:
\begin{align}
\theomsym{\indnuci}{\indnucii}{\ther{}}{B1}
&=- 
\theq{\indnuci}{\indnucii}{\ther{}} \thebt{} ,
\label{cha_010}
\\
\theomsym{\indnuci}{\indnucii}{\ther{}}{B2}
&= - \theq{\indnuci}{\indnucii}{\ther{}} \thebt{}
- \thep{\indnuci}{\indnucii}{\ther{}} \thegas{\indnuci}{\indnucii}{\ther{}}  ,
\label{cha_011}
\end{align}
where $\thegas{\indnuci}{\indnucii}{\ther{}}$ depends only on the orientation of the molecule relative to the magnetic field:
\begin{align}
\thegas{\indnuci}{\indnucii}{\ther{}} &=
\thebt{} \therbar{\indnuci}{\indnucii} \therbar{\indnuci}{\indnucii}^\mathrm{T} -
\therbar{\indnuci}{\indnucii} \therbar{\indnuci}{\indnucii}^\mathrm{T} \thebt{} .
\label{cha_009}
\end{align}
Analyzing the permutational symmetry of the B2 model, we see that $\theq{\indnuci}{\indnucii}{\ther{}}$ and $\thep{\indnuci}{\indnucii}{\ther{}}$ recover the symmetric and antisymmetric parts, respectively:
\begin{align}
\theomsym{\indnuci}{\indnucii}{\ther{}}{B2+}
&=- 
\theq{\indnuci}{\indnucii}{\ther{}} \thebt{} ,
\label{cha_011b}
\\
\theomsym{\indnuci}{\indnucii}{\ther{}}{B2-}
&= 
- \thep{\indnuci}{\indnucii}{\ther{}} \thegas{\indnuci}{\indnucii}{\ther{}}  .
\label{cha_011c}
\end{align}
We can therefore also set up the charge approximations by invoking the decomposition of a matrix into a scalar ($s_{\indnuci\indnucii}$), a vector ($\mathbf{V}_{\indnuci\indnucii}$), and a traceless matrix ($\mathbf{T}_{\indnuci\indnucii}$):\cite{Smith1992}
\begin{align}
\theom{\indnuci}{\indnucii}{\ther{}} &=
s_{\indnuci\indnucii} \, \theone{} +
\mathbf{\tilde{V}}_{\indnuci\indnucii} +
\mathbf{T}_{\indnuci\indnucii}.
\label{cha_012}
\end{align}
In this case, $-\theq{\indnuci}{\indnucii}{\ther{}}$ can be interpreted as the component of $\mathbf{V}_{\indnuci\indnucii}$ along $\theb{}$, whereas $\thep{\indnuci}{\indnucii}{\ther{}} \thegas{\indnuci}{\indnucii}{\ther{}}$ is the component of $\mathbf{T}_{\indnuci\indnucii}$ constructed from the orthogonal axes $\therbar{\indnuci}{\indnucii}$ and $\theb{} \times \therbar{\indnuci}{\indnucii}$.

Note that the B2 model is exact for diatomic molecules oriented perpendicular to the magnetic field.\cite{Culpitt2021} With the magnetic field aligned with the $z$-axis, the Berry curvature then takes the following simple form, which is perfectly captured by the B2 approximation and its two charges:
\begin{align}
&\theom{\indnuci}{\indnucii}{\ther{}} = \theomsym{\indnuci}{\indnucii}{\ther{}}{B2} = \nonumber \\
& |\theb{}| \times \begin{pmatrix}
0 & \theq{\indnuci}{\indnucii}{\ther{}} \!-\! \thep{\indnuci}{\indnucii}{\ther{}} & 0 \\
- \theq{\indnuci}{\indnucii}{\ther{}} \!-\! \thep{\indnuci}{\indnucii}{\ther{}} & 0 & 0 \\
0 & 0 & 0
\end{pmatrix}.
\label{cha_012b}
\end{align}
The $\theq{\indnuci}{\indnucii}{\ther{}}$ charges are thus identical to the screening charges introduced to analyze the Berry curvature in Ref.~\onlinecite{Culpitt2021}. The contribution from the $\thep{\indnuci}{\indnucii}{\ther{}}$ charges in Eq.~\eqref{cha_012} vanishes for H$_2$ or when we assume that the diatomic system is aligned with the magnetic field. For these systems, the B1 model correctly reproduces the behavior of the exact Berry curvature:
\begin{align}
\theom{\indnuci}{\indnucii}{\ther{}} &= \theomsym{\indnuci}{\indnucii}{\ther{}}{B1} 
\nonumber \\ &= |\theb{}| \begin{pmatrix}
0 & \theq{\indnuci}{\indnucii}{\ther{}}  & 0 \\
- \theq{\indnuci}{\indnucii}{\ther{}} & 0 & 0 \\
0 & 0 & 0
\end{pmatrix}.
\label{cha_013}
\end{align}

Even for a general molecule,  the B1 and the B2 models may be useful alternatives to the full Berry curvature, especially when conducting molecular dynamics simulations in magnetic fields. The remaining challenge is, however, to determine the $\theq{\indnuci}{\indnucii}{\ther{}}$ and $\thep{\indnuci}{\indnucii}{\ther{}}$ charges without  calculating the Berry curvature. Such an attempt was made in the M2 model of Ref.~\onlinecite{Peters2022}, where the $\theq{\indnuci}{\indnucii}{\ther{}}$ charges were replaced by Mulliken overlap populations. Here, we  focus on the physical interpretation of the Berry curvature via the charges $\theq{\indnuci}{\indnucii}{\ther{}}$ and $\thep{\indnuci}{\indnucii}{\ther{}}$.

\subsection{Interpretation of Berry Charges and Charge Fluctuations}

\begin{figure}
\begin{tabular}{ll}
(a) & (b) \\
\includegraphics[width=0.23\textwidth, trim=70 150 110 90,clip]{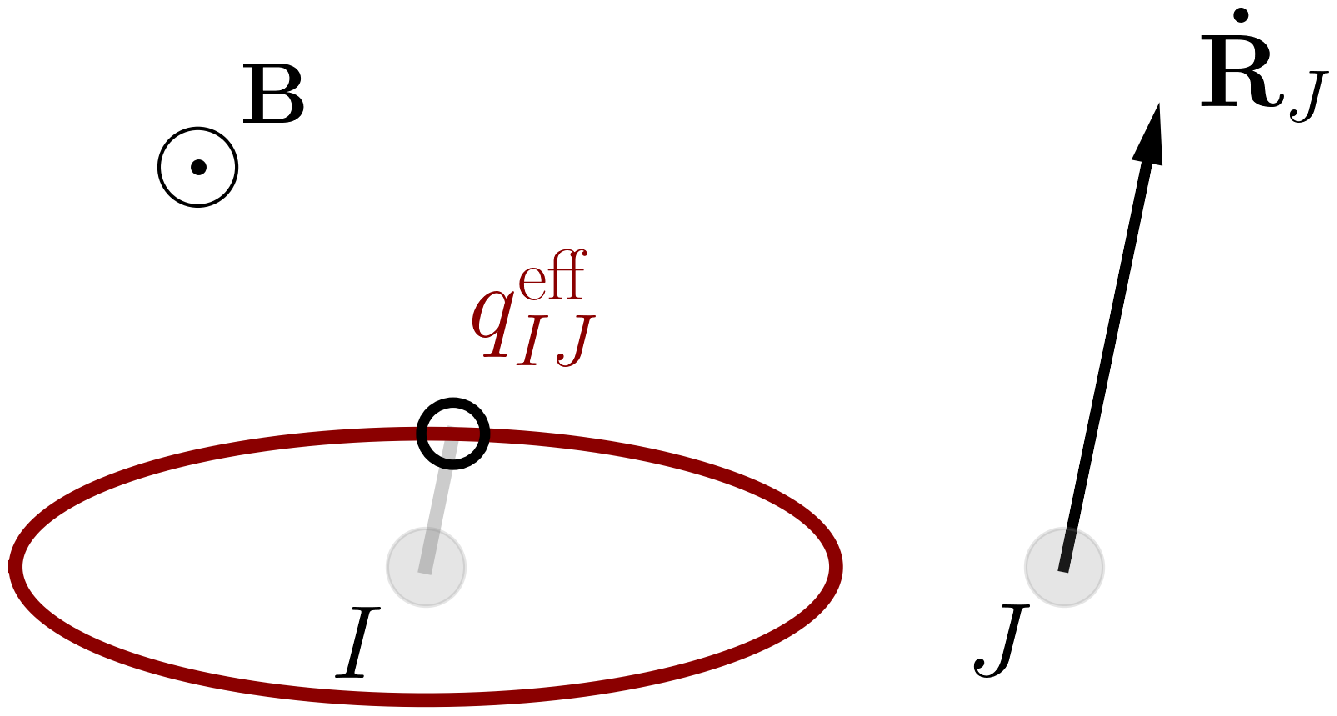} &
\includegraphics[width=0.23\textwidth, trim=70 150 110 90,clip]{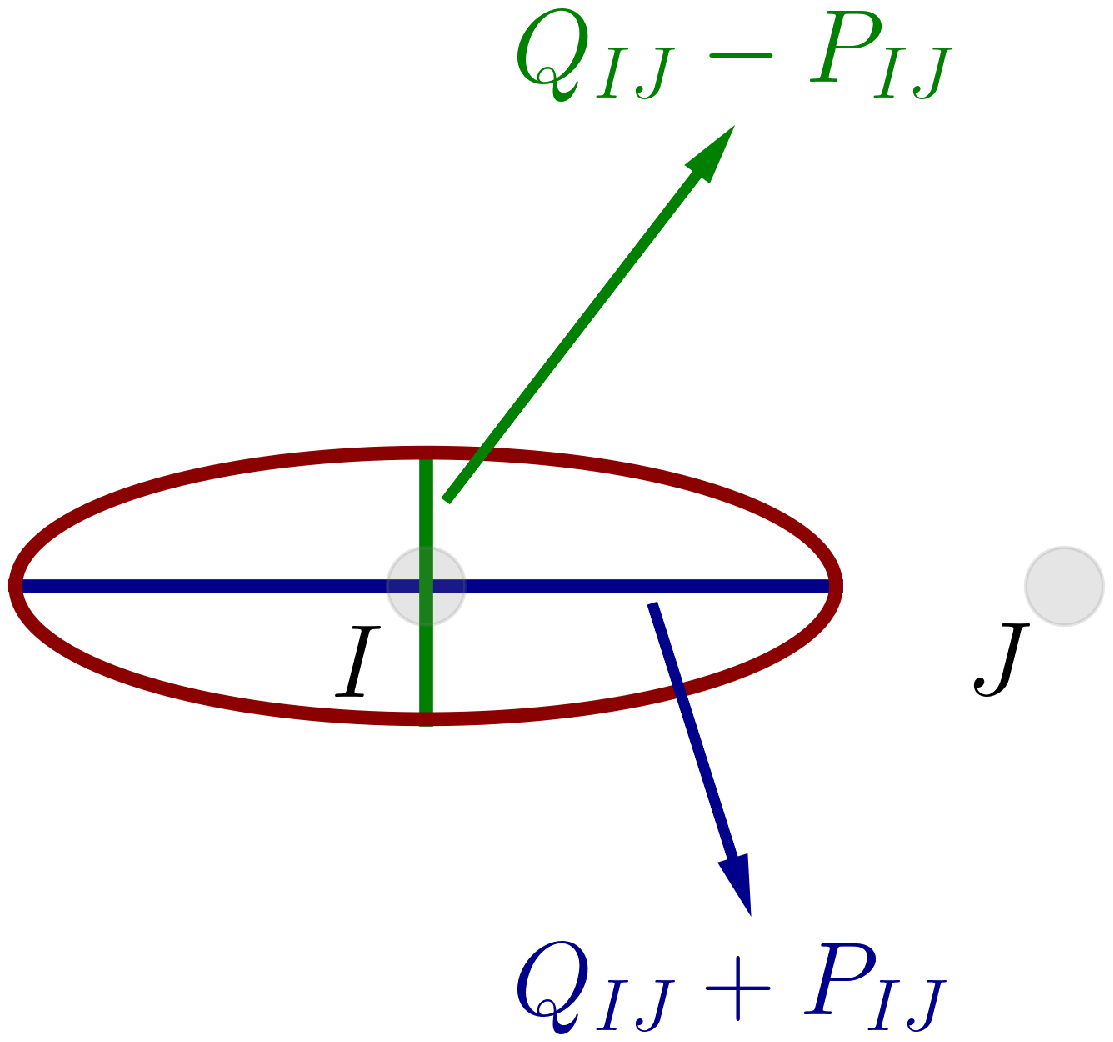} 
\end{tabular}
\caption{Geometric illustration of the directional dependence of the effective charge on $\indnuci$ ($q_{\indnuci\indnucii}^\mathrm{eff}$) from the velocity of $\indnucii$ ($\therdot{\indnucii}$) for a diatomic molecule perpendicular to the magnetic field $\mathbf{B}$. (a) The radius of the ellipse in the direction of $\therdot{\indnucii}$ corresponds to $q_{\indnuci\indnucii}^\mathrm{eff}$. (b) The shape of the ellipse is determined by the Berry charges and charge fluctuations.}
\label{fig_scheme}
\end{figure}

The most straightforward way to understand the role of $\theq{\indnuci}{\indnucii}{\ther{}}$ and $\thep{\indnuci}{\indnucii}{\ther{}}$ follows from the special case of diatomic molecules perpendicular to the field. When the molecule and the magnetic field are aligned with the $x$- and $z$-axes, respectively, the Berry curvature takes the form of Eq.~\eqref{cha_012} and we can write the Cartesian components of the Berry force as
\begin{align}
- \thefdir{\indnuci\indnucii}{B}{x} &= -
|\theb{}| \left[\theq{\indnuci}{\indnucii}{\ther{}} - \thep{\indnuci}{\indnucii}{\ther{}} \right] \therdotdir{\indnucii}{y} ,\label{dia_000}\\
\thefdir{\indnuci\indnucii}{B}{y} &= -
|\theb{}| \left[\theq{\indnuci}{\indnucii}{\ther{}} + \thep{\indnuci}{\indnucii}{\ther{}} \right] \therdotdir{\indnucii}{x} ,\label{dia_001}\\
\thefdir{\indnuci\indnucii}{B}{z} &= 0  , \label{dia_002}
\end{align}
where $\thef{\indnuci\indnucii}{B}$ is the Berry force on atom $\indnuci$ induced by the movement of atom $\indnucii$. Equations~\eqref{dia_000} and \eqref{dia_001} can be interpreted as Lorentz forces with $\theq{\indnuci}{\indnucii}{\ther{}}$ and $\thep{\indnuci}{\indnucii}{\ther{}}$ serving as charges. However, the signs of these charges depend on the direction of the velocity. 

To investigate this further, it is helpful to expand the squared norm of the Berry force as
\begin{align}
|\thef{\indnuci\indnucii}{B}|^2 &= (\thef{\indnuci\indnucii}{B})^\mathrm{T} \thef{\indnuci\indnucii}{B} =
\therdot{\indnucii}^\mathrm{T} \left[ \theomsym{\indnuci}{\indnucii}{\ther{}}{}\right]^\mathrm{T}
\theomsym{\indnuci}{\indnucii}{\ther{}}{} \therdot{\indnucii} \nonumber \\
&= |\theb{}|^2
\therdot{\indnucii}^\mathrm{T} 
\thelamsym{\indnuci}{\indnucii}{\ther{}}{} 
\therdot{\indnucii} ,
\label{dia_003}
\end{align}
where $\thelamsym{\indnuci}{\indnucii}{\ther{}}{} = |\theb{}|^{-2} \left[ \theomsym{\indnuci}{\indnucii}{\ther{}}{}\right]^\mathrm{T}
\theomsym{\indnuci}{\indnucii}{\ther{}}{}$ is a symmetric, positive semidefinite matrix. It can therefore be diagonalized,
\begin{align}
\thelamsym{\indnuci}{\indnucii}{\ther{}}{} =
\sum \limits_{\alpha=1}^3 \lambda_{\indnuci\indnucii}^\alpha \mathbf{v}_{\indnuci\indnucii}^\alpha \mathbf{v}_{\indnuci\indnucii}^{\alpha\mathrm{T}} ,
\label{dia_005}
\end{align}
and the eigenvectors $\mathbf{v}_{\indnuci\indnucii}^\alpha$ may be thought of as the directions of the semiaxes of an ellipsoid. The eigenvalues satisfy $\lambda_{\indnuci\indnucii}^\alpha \geq 0$ and $\sqrt{\lambda_{\indnuci\indnucii}^\alpha}$ are the lengths of the semiaxes. Hence, we may visualize the effect of $\theomsym{\indnuci}{\indnucii}{\ther{}}{}$ on a velocity $\therdot{\indnucii}$ by plotting the ellipsoid:
\begin{equation}
   \therdot{\indnucii}^\mathrm{T} \, \thelamsym{\indnuci}{\indnucii}{\ther{}}{}^{-1} \, \therdot{\indnucii} = \mathrm{const}.
   \label{eq_elipse}
\end{equation}
Since one eigenvalue of $\thelamsym{\indnuci}{\indnucii}{\ther{}}{}$ is zero, the ellipsoid is a flat disk and we interpret $\thelamsym{\indnuci}{\indnucii}{\ther{}}{}^{-1}$ as a generalized inverse. In the context of the Berry force and in line with our starting point in eq.~\eqref{cur_007}, $\lambda_{\indnuci\indnucii}^\alpha$ can be interpreted as the square of an \emph{effective charge} $\thecharge{\indnuci\indnucii}{eff,\alpha}{\ther{}}$ that weights the Lorentz force due to a velocity $\therdot{\indnucii}$ along $\mathbf{v}_{\indnuci\indnucii}^\alpha$.

In the special case of a linear molecule perpendicular to the field, $\thelamsym{\indnuci}{\indnucii}{\ther{}}{}$ is a diagonal matrix from which we can directly determine the eigenvalues as 
\begin{align}
\left[\thecharge{\indnuci\indnucii}{eff}{\ther{}}\right]^2 &=
\left[\theq{\indnuci}{\indnucii}{\ther{}} \pm \thep{\indnuci}{\indnucii}{\ther{}} \right]^2 .
\label{dia_007}
\end{align}
The eigenvalue along the $z$-axis is zero, so that our ellipsoid of eq.~\eqref{eq_elipse} collapses to an ellipse with $\theq{\indnuci}{\indnucii}{\ther{}} \pm \thep{\indnuci}{\indnucii}{\ther{}}$ as principle axes along the $x$- and $y$-directions, respectively; see Fig.~\ref{fig_scheme}. From this simple picture, we see that $\theq{\indnuci}{\indnucii}{\ther{}}$ is the \emph{isotropic} component of the charge, being independent of the direction of the velocity. This observation agrees with our conclusion in Ref.~\onlinecite{Culpitt2021} -- namely, that $\theq{\indnuci}{\indnucii}{\ther{}}$ represents the amount of electrons by which nucleus $\indnucii$ screens nucleus $\indnuci$ or \emph{vice versa}. For this reason, we refer to the $\theq{\indnuci}{\indnucii}{\ther{}}$ as the Berry charges from now on. The $\thep{\indnuci}{\indnucii}{\ther{}}$ can be understood as a measure of \emph{anisotropy} -- that is, the amount by which the effective charge fluctuates with direction of the velocity. We will therefore
refer to the $\thep{\indnuci}{\indnucii}{\ther{}}$ as Berry charge fluctuations.

Let us now consider a molecule with an arbitrary angle $\theta$ between $\therbar{\indnuci}{\indnucii}$ and $\theb{}$. In the appendix, we show that, using the B2 model for the Berry curvature, we obtain the following two non-zero eigenvalues:
\begin{align}
\left[\thecharge{\indnuci\indnucii}{eff}{\ther{}}\right]^2 &=\left[\theq{\indnuci}{\indnucii}{\ther{}}\right]^2
+ \sin^2(\theta)\left[\thep{\indnuci}{\indnucii}{\ther{}}\right]^2
\nonumber \\
&\quad\pm 
2 \theq{\indnuci}{\indnucii}{\ther{}} \thep{\indnuci}{\indnucii}{\ther{}}
\sin^2(\theta)
\label{dia_008}
\end{align}
This demonstrates that the contribution of the Berry charge fluctuations depends on $\theta$. For a molecule parallel to the magnetic field, the effective charge becomes isotropic, while an angle of $\pi/2$ reproduces the result in Eq.~\eqref{dia_007}:
\begin{equation}
\left[\thecharge{\indnuci\indnucii}{eff}{\ther{}}\right]^2 = \begin{cases}
\left[\theq{\indnuci}{\indnucii}{\ther{}}\right]^2,&\theta = 0 , \\
\left[\theq{\indnuci}{\indnucii}{\ther{}} \pm \thep{\indnuci}{\indnucii}{\ther{}} \right]^2,&\theta = \pi/2.
\end{cases}
\end{equation}
In the latter case, the eigenvectors are aligned with $\therbar{\indnuci}{\indnucii}$ and $\theb{} \times \therbar{\indnuci}{\indnucii}$, respectively. It should be noted that the exact $\thelamsym{\indnuci}{\indnucii}{\ther{}}{}$ may have different semiaxes than assumed in the B2 approximation. 

\subsection{Berry Population Analysis}

As a final step in this section, we introduce a new population analysis, which we refer to as Berry population analysis. In general, such an analysis aims at extracting atomic charges from quantum-chemical calculations, yielding insights into bonding situations, reactivity, and so on. Prominent examples are the Mulliken charges\cite{Mulliken1955} [$\thecharge{\indnuci}{M}{\ther{}}$] and overlap populations [$\theqm{\indnuci}{\indnucii}{\ther{}}$], which use molecular orbitals constructed only from basis functions assigned to atom $\indnuci$ [$\themomul{\indocci}{\indnuci}{0} (\ther{})$]: 
\begin{align}
\thecharge{\indnuci}{M}{\ther{}} &= \theelcharge{}\thez{\indnuci} + \sum \limits_{\indnucii=1}^{\then{nuc}}
\theqm{\indnuci}{\indnucii}{\ther{}},
\label{bpa_000b} \\
\theqm{\indnuci}{\indnucii}{\ther{}}
&= -
\dfrac{1}{2}
\sum \limits_{\indocci=1}^{\then{occ}} 
\bigg[
\Braket{
\themomul{\indocci}{\indnucii}{0} (\ther{}) |
\themomul{\indocci}{\indnuci}{0} (\ther{})
}
\nonumber \\
&\qquad\qquad \quad+
\Braket{
\themomul{\indocci}{\indnuci}{0} (\ther{})|
\themomul{\indocci}{\indnucii}{0} (\ther{}) 
}
\bigg],
\label{bpa_001b}
\end{align}
and the generalized atomic-polar-tensor charges\cite{Cioslowski1989} [$\thecharge{\indnuci}{D}{\ther{}}$], which determine the isotropic part of the atomic polar tensor:
\begin{align}
\thecharge{\indnuci}{D}{\ther{}} &= \theelcharge{}\thez{\indnuci} + \dfrac{1}{3} \mathrm{Tr} \big[ \theapt{\indnuci}{\ther{}}{0} \big].
\label{bpa_000}
\end{align}

In the previous subsections, we have shown that the Berry curvature can indeed be interpreted as an effective charge. However, since the effective charge depends on the direction of the velocity, there is no unique definition of an atomic charge. Here, we define the \emph{Berry atomic charge} [$\thecharge{\indnuci}{B}{\ther{}}$] as the sum over all isotropic Berry charges and the nuclear charge:
\begin{align}
\thecharge{\indnuci}{B}{\ther{}} &= \theelcharge{} \thez{\indnuci} + \sum \limits_{\indnucii=1}^{\then{nuc}}
\theq{\indnuci}{\indnucii}{\ther{}} .
\label{bpa_001}
\end{align}
This choice ensures that the charge is isotropic and corresponds to the effective charge on $\indnuci$ during a rigid translation of the entire molecule $\therdot{\mathrm{T}}$:
\begin{align}
\thef{\indnuci}{L}  + \thef{\indnuci}{B1}  &=
- \thecharge{\indnuci}{B}{\ther{}} \thebt{} \therdot{\mathrm{T}} 
\label{bla_003}
\end{align}

We note that the Berry population analysis requires the presence of a magnetic field. This should be kept in mind when comparing to population analyses performed at zero field.  Additionally, it should be mentioned that the electron density depends on the orientation of the molecule with respect to the magnetic field vector. To account for this indirect dependence, every population analysis in a non-zero field requires rotational averaging $\Braket{\thecharge{\indnuci}{B}{\ther{}}}_\mathrm{rot}$ before evaluation.

\section{Computational Details}

All calculations presented here were performed at the Hartree--Fock/l-cc-pVDZ level of theory using the {\sc London}\cite{London} program package, at the zero-field optimized molecular geometries. Here l-cc-pVDZ denotes the London-orbital variant of the contracted cc-pVDZ\cite{Dunning1989} basis set, which has been shown to give Berry curvatures in good agreement with results obtained from the computationally more expensive l-cc-pVTZ basis set.\cite{Peters2022} The magnetic field strength was set to $0.001\theb{0}$. The Berry curvature and the atomic-polar-tensor charges were obtained from finite differences calculations with a step size of $5 \times 10^{-4}\,$Bohr. The error of this numerical approach is less than $0.01\%$.\cite{Peters2022} The Berry charges and charge fluctuations were subsequently obtained by solving Eqs.~\eqref{cha_011b} and \eqref{cha_011c} in a least-squares fashion:
\begin{align}
\theq{\indnuci}{\indnucii}{\ther{}}
&=- 
\dfrac{\mathrm{Tr}\left\{\left[\theomsym{\indnuci}{\indnucii}{\ther{}}{+}\right]^\mathrm{T}\thebt{} \right\}}
{\mathrm{Tr}\left\{\thebt{} ^\mathrm{T}\thebt{} \right\} },
\label{lsq_000}
\\
\thep{\indnuci}{\indnucii}{\ther{}}
&=- 
\dfrac{\mathrm{Tr}\left\{\left[\theomsym{\indnuci}{\indnucii}{\ther{}}{-}\right]^\mathrm{T}\thegas{\indnuci}{\indnucii}{\ther{}} \right\}}
{\mathrm{Tr}\left\{\left[\thegas{\indnuci}{\indnucii}{\ther{}} \right]^\mathrm{T}\thegas{\indnuci}{\indnucii}{\ther{}}  \right\} }.
\label{lsq_001}
\end{align}

From these quantities, we calculate the Berry atomic charges $\thecharge{\indnuci}{B}{\ther{}}$ and the corresponding approximate Berry curvatures $\theomsym{\indnuci}{\indnucii}{\ther{}}{B1}$ and $\theomsym{\indnuci}{\indnucii}{\ther{}}{B2}$. The Mulliken charges $\thecharge{\indnuci}{M}{\ther{}}$, overlap populations $\theqm{\indnuci}{\indnucii}{\ther{}}$, and Mulliken approximation to the Berry curvature $\theomsym{\indnuci}{\indnucii}{\ther{}}{M2}$ [see ref.~\onlinecite{Peters2022}] were obtained from single-point calculations. As in Ref.~\onlinecite{Peters2022}, we quantify the error of an approximate Berry curvature by the following \enquote{screening error per electron}:
\begin{align}
\epsilon_\mathrm{X} = 
\dfrac{
\big|\big| 
\theomsym{}{}{\ther{}}{X}  - \theom{}{}{\ther{}} \big|\big|_2
}{
\big|\big| 
\theom{}{}{\ther{}} \big|\big|_2
},
\label{error}
\end{align}
For the B2 approximation, we also determine $\epsilon_\mathrm{S}^\mathrm{B2}$, $\epsilon_\mathrm{V}^\mathrm{B2}$, and $\epsilon_\mathrm{T}^\mathrm{B2}$ as the errors within the scalar, vector, and tensor parts of the Berry curvature [see Eq.~\eqref{cha_012}]:
\begin{align}
\left[\epsilon_\mathrm{B2}\right]^2 &= \left[\epsilon_\mathrm{S}^\mathrm{B2} \right]^2 +  \left[\epsilon_\mathrm{V}^\mathrm{B2} \right]^2 +  \left[\epsilon_\mathrm{T}^\mathrm{B2} \right]^2
\label{error2}
\end{align}

For the rotationally averaged values $\braket{\thecharge{\indnuci}{X}{\ther{}}}_\mathrm{rot}$, $\Braket{\theq{\indnuci}{\indnucii}{\ther{}}}_\mathrm{rot}$, and $\braket{\epsilon_\mathrm{X}}_\mathrm{rot}$, we performed a numerical spherical integration over 146 geometries with different orientations relative to the magnetic field, using points and weights from the {\sc Python3.6} {\sc quadpy}\cite{quadpy} package. As shown in Fig.~S1 in the Supporting Information, a grid of 146 points is sufficient to calculate the rotational average with an error below $0.1\%$ for the current setup. The magnetic field is small enough ($0.001\theb{0}$) that we can use the same structure for all orientations.

\section{Results and Discussion}

\subsection{Validation of the Charge Approximations}

\begin{figure*}
\begin{tabular}{ll}
(a) & \\
\multicolumn{2}{c}{\includegraphics[width=0.98\textwidth]{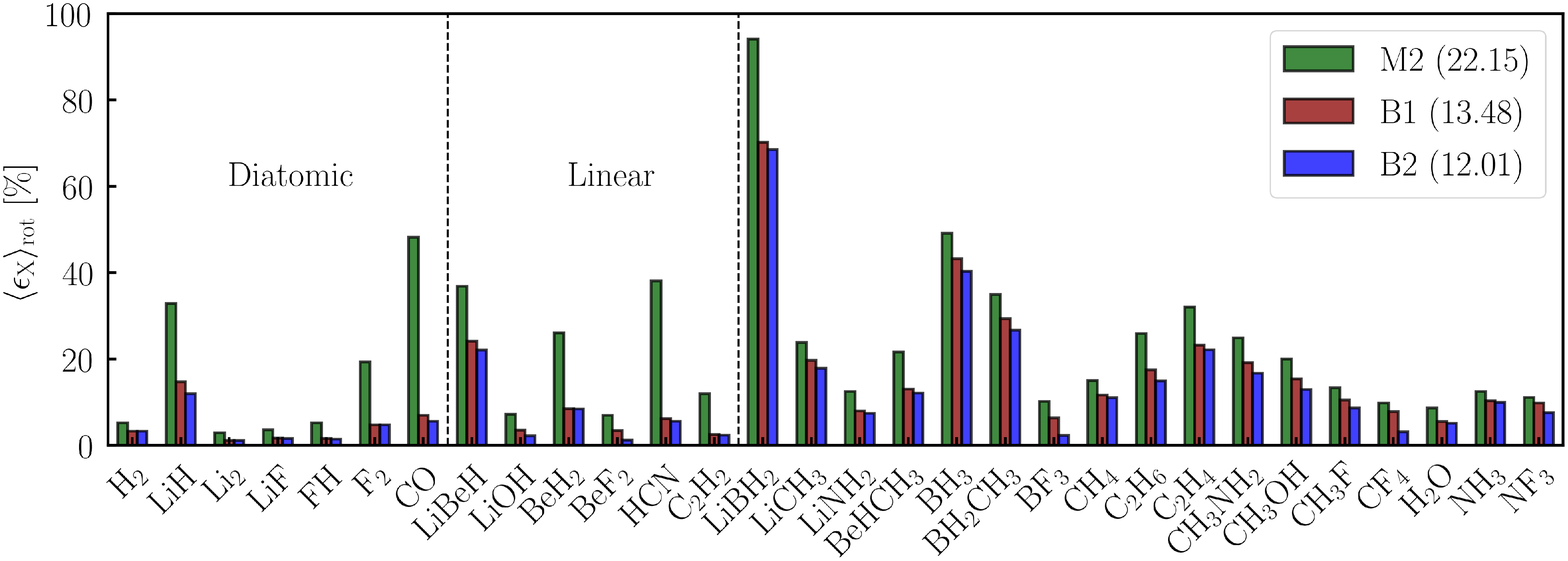}}  \\
(b) & \\
\multicolumn{2}{c}{\includegraphics[width=0.98\textwidth]{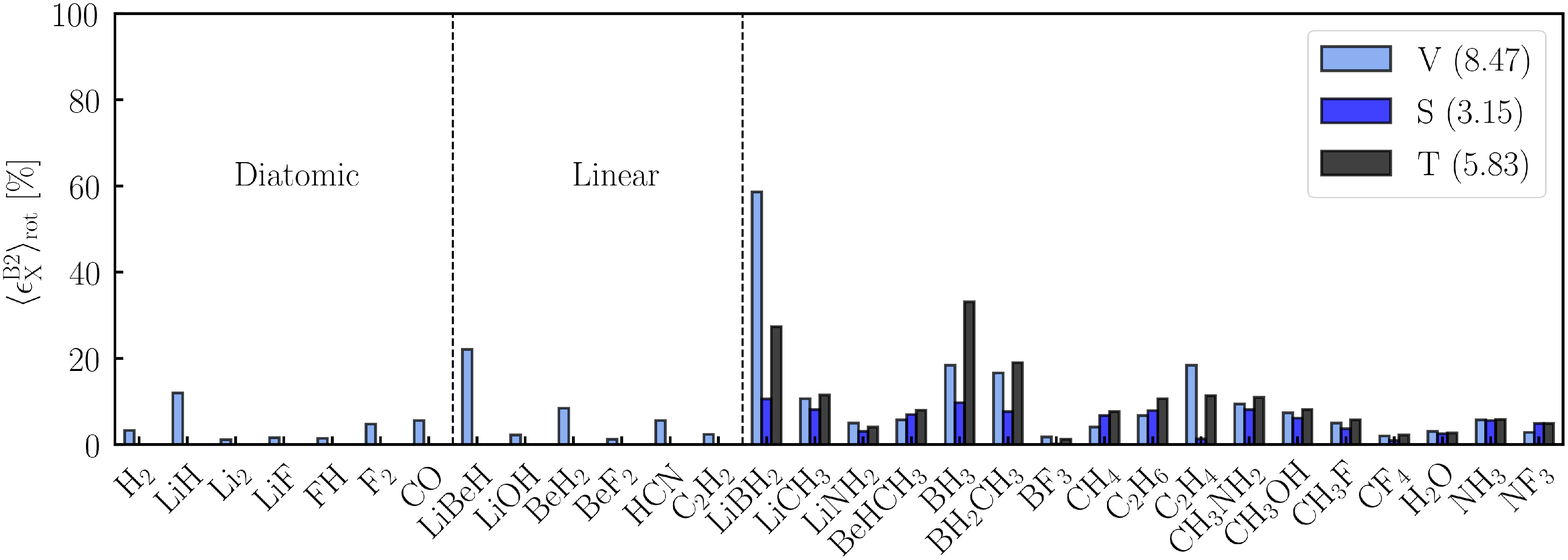}}  \\
(c) & (d) \\
\includegraphics[width=0.48\textwidth]{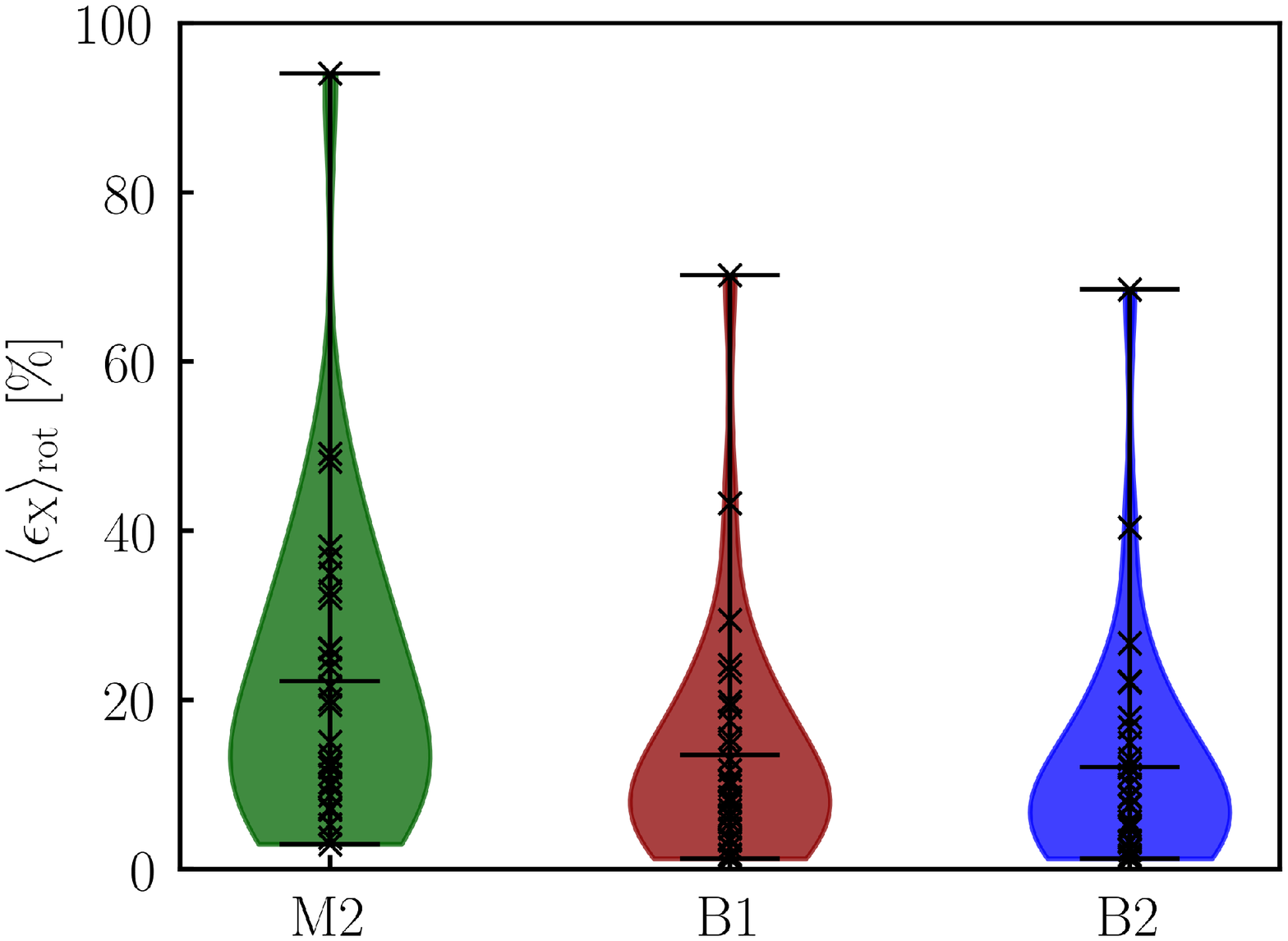}  &
\includegraphics[width=0.48\textwidth]{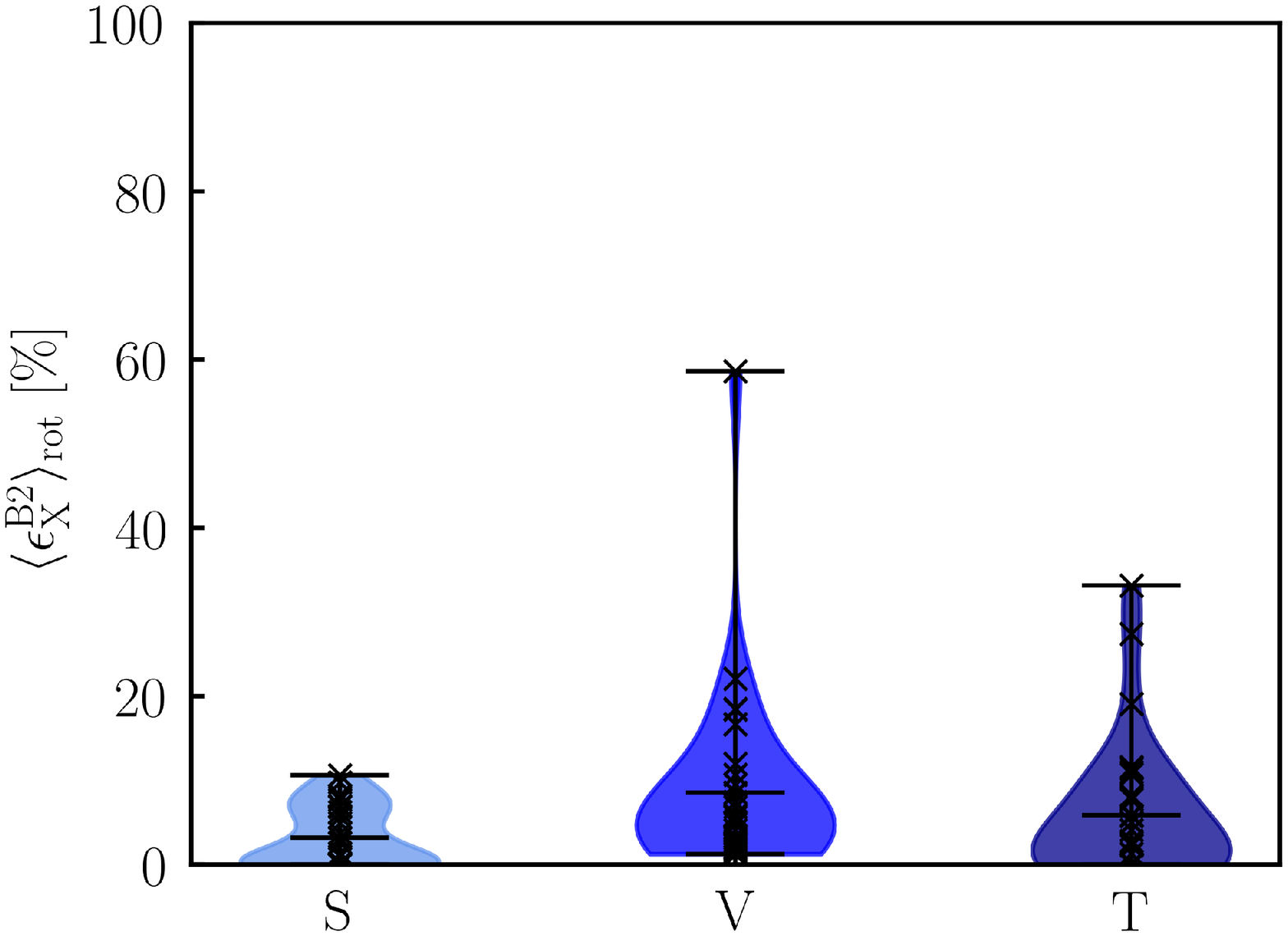}  
\end{tabular}
\caption{Rotationally averaged errors [see Eq.~\eqref{error}] of (a) the approximate Berry curvatures M2 [Ref.~\onlinecite{Peters2022}], B1 [Eq.~\eqref{cha_010}], and B2 [Eq.~\eqref{cha_011}] and (b) the scalar (S), vector (V), and tensor (T) components of the B2 approximation calculated for a series of molecules. The mean error of all molecules is given in brackets. In (c) and (d), we show violin plots for all the data points in (a) and (b), respectively.
}
\label{fig_approx}
\end{figure*}

Before discussing the Berry charges, charge fluctuations, and corresponding population analyses, we need to validate the charge approximations B1 [see eq.~\eqref{cha_010}] and B2 [see eq.~\eqref{cha_011}]. The main question is how much of the exact Berry curvature is captured by $\theq{\indnuci}{\indnucii}{\ther{}}$ and $\thep{\indnuci}{\indnucii}{\ther{}}$. As a measure of the screening error, we calculate the rotational average of $\epsilon_\mathrm{B1}$ and $\epsilon_\mathrm{B2}$ for a set of 30 molecules -- see Fig.~\ref{fig_approx}, where we also show the error of the second Mulliken approximation [M2, see Ref.~\onlinecite{Peters2022}]  for comparison as well as a decomposition of $\epsilon_\mathrm{B2}$ into contributions arising from the scalar, vector, and tensor component of the Berry curvature [see eq.~\eqref{error2}].

The new approximations perform significantly better than the Mulliken approximation; see Fig.~\ref{fig_approx}(a). For every molecule, B1 is closer to the exact Berry curvature, the average error decreasing from 22.15\% for M2 to 13.48\% for B1. This reduction is expected since the $\theq{\indnuci}{\indnucii}{\ther{}}$ charges in the B1 model are chosen to minimize $\epsilon$. Inclusion of the Berry charge fluctuations $\thep{\indnuci}{\indnucii}{\ther{}}$ reduces the error further to 12.01\% for the B2 model. As expected from our derivation of the B2 model, this effect is especially strong for linear molecules, where the errors from the scalar and tensor contribution vanish. For nonlinear molecules, the error appears to be evenly spread among the different contributions [S, V, T in Fig.~\ref{fig_approx}(b)], indicating that it would require at least three additional parameters (or charges) to significantly reduce the error. 

The violin plots in Fig.~\ref{fig_approx}(c+d) show that, even for the B2 approximation, there are several outliers -- for example, LiH, BH$_3$, LiBeH, and LiBH$_2$. As  discussed in Ref.~\onlinecite{Peters2022}, our charge models are probably less accurate for low-valent molecules. In general, we  conclude that many features of the Berry curvature are well captured by the proposed Berry charges and charge fluctuations, which will therefore be a useful tool to study the effect of $\theom{}{}{\ther{}}$ on molecular dynamics.

\subsection{Berry Charges and Charge Fluctuations}

\begin{figure}
\begin{tabular}{llll}
(a) H$_2$ & (b) LiH \\
\includegraphics[width=0.23\textwidth, trim=50 90 50 90,clip]{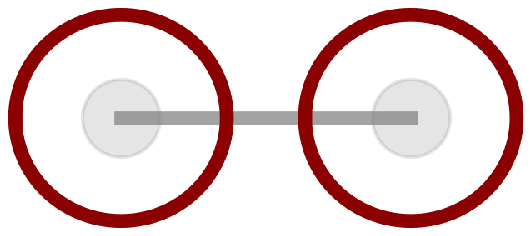} &
\includegraphics[width=0.23\textwidth, trim=50 90 50 90,clip]{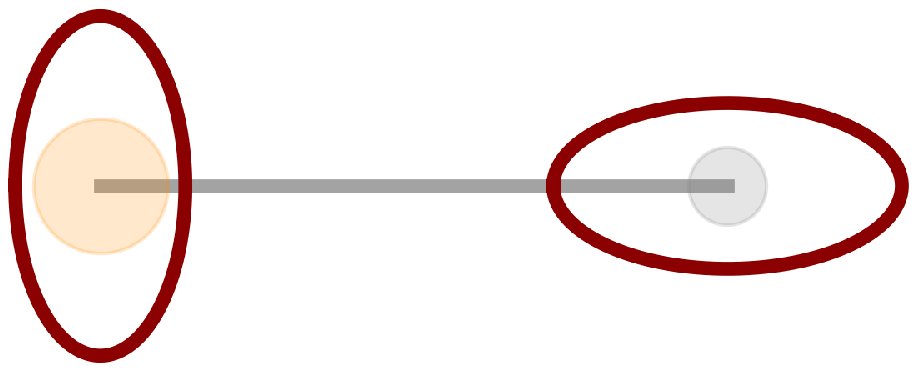} \\
(c) FH & (d) HCN \\
\includegraphics[width=0.23\textwidth, trim=50 90 50 90,clip]{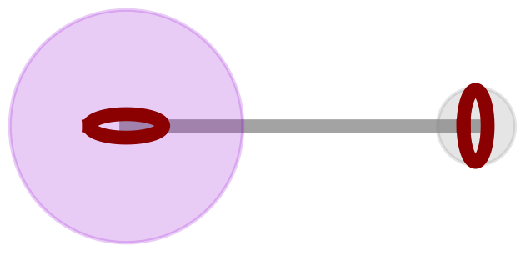} &
\includegraphics[width=0.23\textwidth, trim=50 90 50 90,clip]{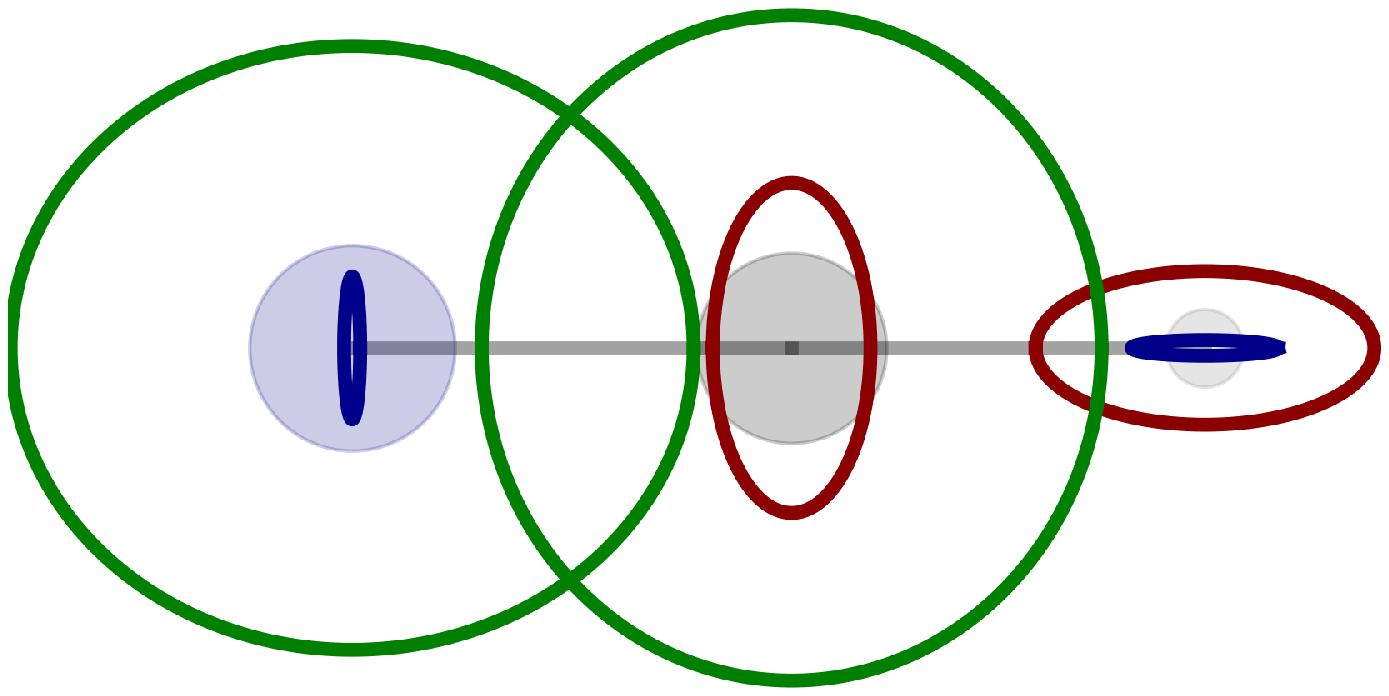} \\
(e) BH$_3$ & (f) H$_2$O \\
\includegraphics[width=0.23\textwidth, trim=50 90 50 90,clip]{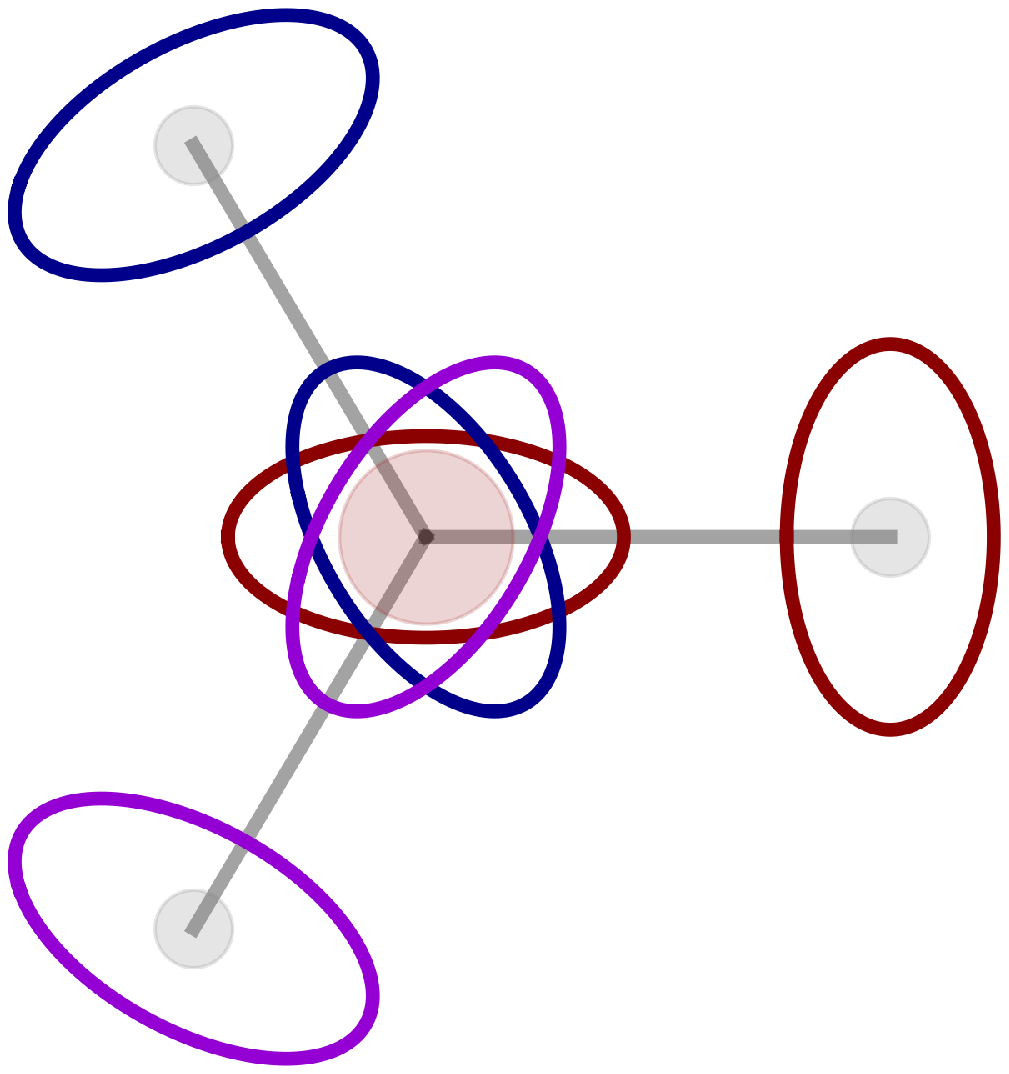} &
\includegraphics[width=0.23\textwidth, trim=50 90 50 90,clip]{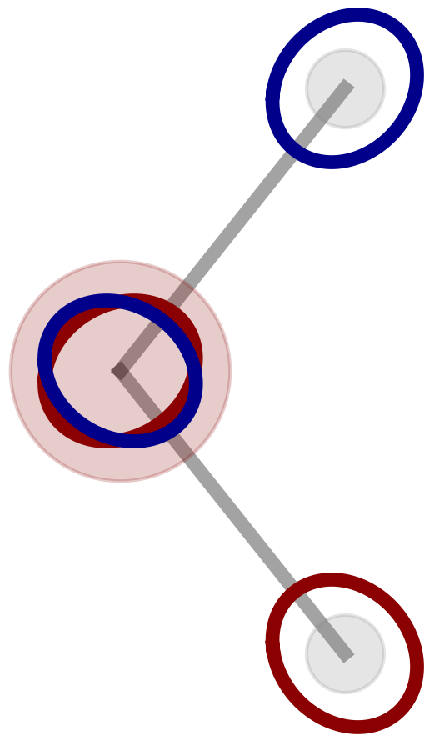} 
\end{tabular}
\caption{Visualization of the effective charges $q_{\indnuci\indnucii}^\mathrm{eff}$, obtained from the Berry charges and charge fluctuations, for a series of planar molecules perpendicular to the magnetic field.
Ellipses representing different pairs $\indnuci\indnucii$ are shown in a different colors and the ellipse representing $\indnuci\indnucii$ is centered at $\mathbf{R}_{\indnuci}$. For brevity, we do not show the $q_{\indnuci\indnuci}^\mathrm{eff}$ in (a)--(f) and the $q_{\mathrm{H}\mathrm{H}}^\mathrm{eff}$ in (e)--(f).}
\label{fig_mols}
\end{figure}

To investigate $\theq{\indnuci}{\indnucii}{\ther{}}$ and $\thep{\indnuci}{\indnucii}{\ther{}}$, we consider a test set of eight small molecules (H$_2$, LiH, BH$_3$, CH$_4$, HCN, NH$_3$, H$_2$O, and FH), representing a wide range of bond types, electronegativity differences, and orientations towards the magnetic field vector; see Tables~\ref{tab_charges1} and \ref{tab_charges2}. For the planar molecules in a perpendicular field orientation, the Berry charges are visualized in Fig.~\ref{fig_mols} as discussed in Section~II.D.

\begin{table}
\centering
\caption{Berry charges [$\theq{\indnuci}{\indnucii}{\ther{}}$] and charge fluctuations [$\thep{\indnuci}{\indnucii}{\ther{}}$] of the linear molecules H$_2$, LiH, FH, and HCN in parallel and perpendicular field orientations. For HCN, HX is the sum of the HC and HN contributions.}
\begin{tabular}{cc|ccc}
\hline\hline
Mol. & $\theb{}$ & $\theq{\mathrm{H}}{\mathrm{H}}{\ther{}}$ & $\theq{\mathrm{H}}{\mathrm{X}}{\ther{}}$ & $\thep{\mathrm{H}}{\mathrm{X}}{\ther{}}$ \\ \hline
\multirow{2}{*}{H$_2$} & $\perp$ & $-$0.610 & $-$0.390 & 0.000 \\
 & $||$ & $-$0.559 & $-$0.441 & 0.000 \\ \hline
\multirow{2}{*}{LiH} & $\perp$ & $-$1.073 & $-$0.478 & $-$0.164 \\
 & $||$ & $-$1.674 & $-$0.042 & 0.000 \\ \hline
\multirow{2}{*}{FH} & $\perp$ & $-$0.517 & $-$0.090 & 0.046 \\
 & $||$ & $-$0.271 & $-$0.290 & 0.000 \\ \hline
\multirow{2}{*}{HCN} & $\perp$ & $-$0.387 & $-$0.307 & $-$0.046 \\
 & $||$ & $-$0.570 & $-$0.189 & 0.000 \\
 \hline\hline
\end{tabular}
\label{tab_charges1}
\end{table}

\begin{table}
\centering
\caption{Berry charges [$\theq{\indnuci}{\indnucii}{\ther{}}$] and charge fluctuations [$\thep{\indnuci}{\indnucii}{\ther{}}$] of  molecules of type XH$_n$ (X = B, C, N, O) with different orientations of the principal axis relative to the magnetic field. XH denotes the sum over all hydrogens of the molecule.}

\begin{tabular}{cc|ccc}
\hline\hline
Mol. & $\theb{}$ & $\theq{\mathrm{X}}{\mathrm{X}}{\ther{}}$ & $\theq{\mathrm{X}}{\mathrm{H}}{\ther{}}$ & $\thep{\mathrm{X}}{\mathrm{H}}{\ther{}}$ \\ \hline
\multirow{3}{*}{BH$_3$} & $||$ & $-$2.683 & $-$1.667 & $-$0.522 \\
 & $\perp$ & $-$3.217 & $-$1.294 & $-$1.325 \\ 
 & $\perp$ & $-$3.217 & $-$1.294 & $-$0.884  \\  \hline
 \multirow{3}{*}{CH$_4$} & $||$ & $-$4.145 & $-$1.800 & $-$0.224 \\
 & $\perp$ & $-$4.145 & $-$1.800 & $-$0.296 \\ 
  & $\perp$ & $-$4.145 & $-$1.800 & $-$0.301 \\ \hline
  \multirow{3}{*}{NH$_3$} & $||$ & $-$6.012 & $-$1.212 & 0.057 \\
 & $\perp$ & $-$6.474 & $-$0.832 & $-$0.156  \\ 
 & $\perp$ & $-$6.474 & $-$0.832 & $-$0.293  \\ \hline
  \multirow{3}{*}{H$_2$O} & $||$ & $-$8.168 & $-$0.287 & $-$0.393 \\
 & $\perp$ & $-$7.948 & $-$0.561 & $-$0.206  \\ 
 & $\perp$ & $-$7.825 & $-$0.543 & 0.051  \\ 
 \hline\hline
\end{tabular}
\label{tab_charges2}
\end{table}

We draw two main conclusions. First, the Berry charges and charge fluctuations depend strongly on the field orientation with CH$_4$ being an exception due to its tetrahedral symmetry. Second, the Berry charges correlate with bonding properties of the corresponding molecule. The magnitude of $\theq{\mathrm{H}}{\mathrm{X}}{\ther{}}$ is larger when the HX bond is covalent (less than $-0.4$ for H$_2$ and CH$_4$) and smaller when it is considered to be more ionic (greater than $-0.4$ for H$_2$O and NH$_3$). 
Electronegativity also has an impact: as H becomes more electropositive,  $\theq{\mathrm{H}}{\mathrm{H}}{\ther{}}$  increases as seen for LiH ($-$1.1/$-$1.7 in the perpendicular/parallel field orientation), H$_2$ ($-$0.6/$-$0.6), and FH ($-$0.5/$-$0.3). 

The Berry charge fluctuations are smaller than the Berry charges and may be positive and negative. Charge anisotropy occurs for every molecule except  H$_2$ in Fig.~\ref{fig_mols} -- the largest anisotropies are observed for LiH, BH$_3$, and HCN, where the shapes around the hydrogen atom have the largest deviation from a circle. 

\subsection{Berry Population Analysis}

\begin{figure*}[h]
\begin{tabular}{ll}
(a) Li-Series & (b) H-Series \\
\includegraphics[width=0.48\textwidth]{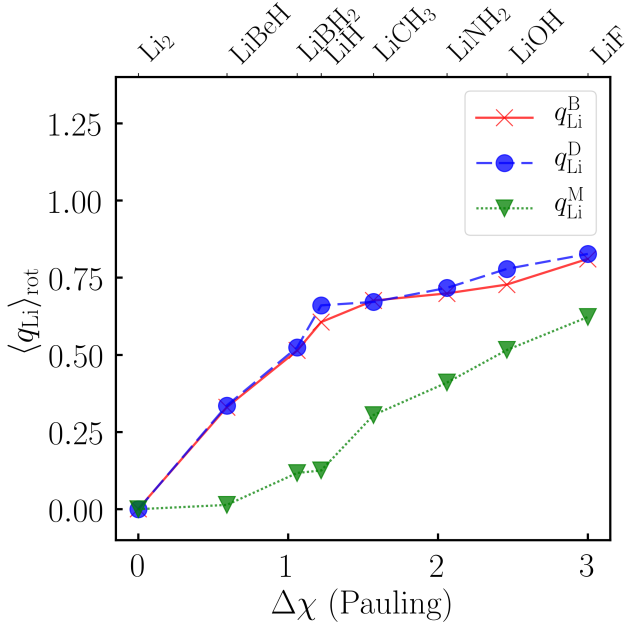} &
\includegraphics[width=0.48\textwidth]{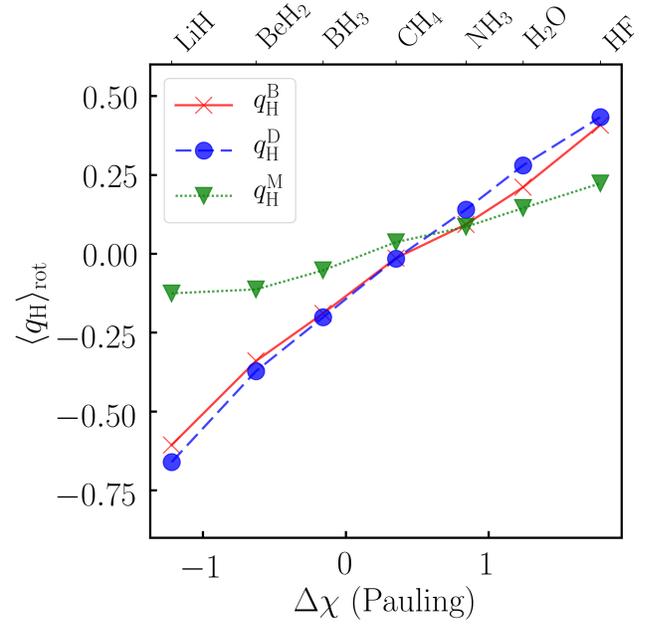} \\
(c) C-Series & (d) F-Series \\
\includegraphics[width=0.48\textwidth]{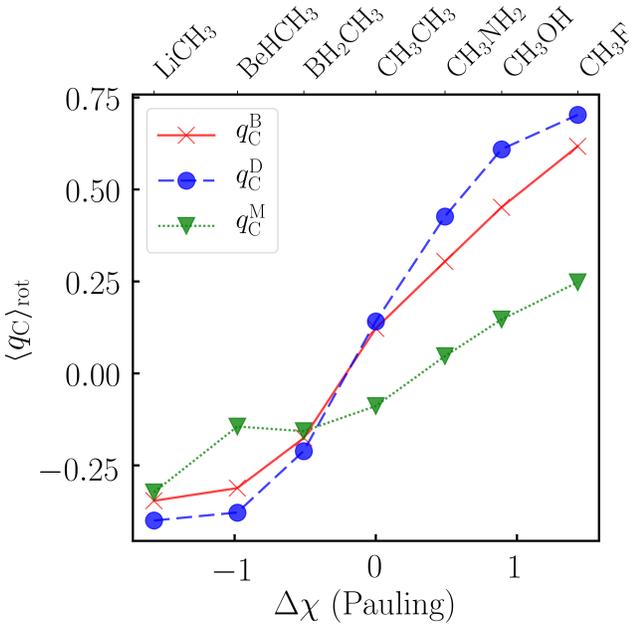} &
\includegraphics[width=0.48\textwidth]{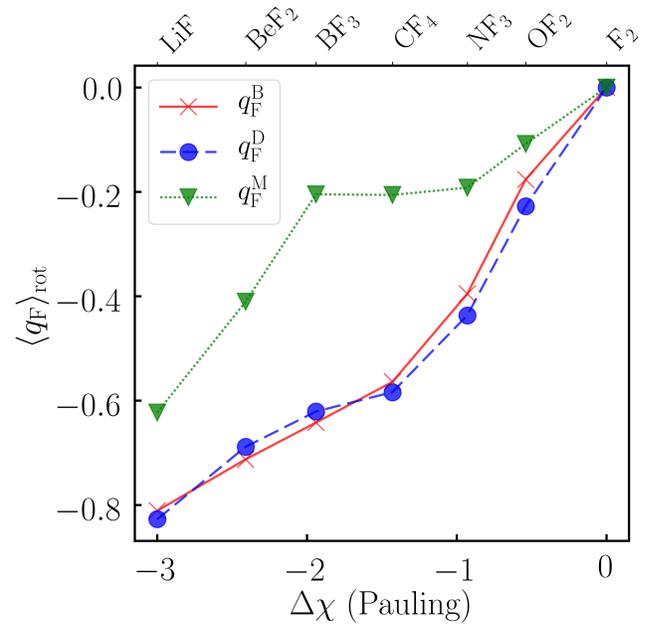} 
\end{tabular}
\caption{Comparison of rotationally averaged Berry (B), atomic-polar-tensor (D), and Mulliken (M) atomic charges ($\smash{q_I^\mathrm{B/D/M}}$, in $\theelcharge{}$, see eqs.~\eqref{bpa_000b}, \eqref{bpa_000}, and \eqref{bpa_001}) for a series of Li- (a), H- (b), C- (c), and F-containing (d) molecules. For each series, the molecules are sorted according to the Pauling electronegativity difference ($\Delta \chi$) between the investigated atom (Li, H, C, and F) and the atom it is bound to. When a molecule consists of more than one atom of the investigated type, we display their averaged charge. The individual values are listed in the Supporting Information.}
\label{fig_comp}
\end{figure*}

A set of Li-, H-, C-, and F-containing molecules with varying electronegativity differences is used to test and validate the Berry population analysis (BPA). As references, the Mulliken\cite{Mulliken1955} (MPA) and the atomic-polar-tensor or dipole population analysis\cite{Cioslowski1989} (DPA) was used, since these methods can easily be adapted to molecules in a magnetic field. For every molecule, we calculate the rotational average of the BPA charges [$\thecharge{\indnuci}{B}{\ther{}}$], MPA charges [$\thecharge{\indnuci}{M}{\ther{}}$], and DPA charges [$\thecharge{\indnuci}{D}{\ther{}}$]. The results of each series are in Fig.\,\ref{fig_comp} plotted against the corresponding Pauling electronegativity\cite{Pauling1932} differences ($\Delta \chi$).

We begin with a brief comparison of the two established population analyses. In all four series, the behavior of the DPA is significantly different from the behavior of the MPA. The DPA covers a wider range of charges and increases monotonically with $\Delta \chi$, while the MPA curves are more bumpy. These results are not surprising given the conceptual differences between the approaches. While we do not consider DPA and MPA methods to yield \enquote{exact} charges at our chosen level of theory (the MPA charges are notorious for their basis-set dependence and the DPA charges suffer from erratic dipole moments at the HF level of theory), the DPA charges appear to be more natural in our test cases.

Interestingly, the BPA charges are in good agreement with the DPA charges in all four series, but differ significantly from the MPA charges. The average difference between the BPA and DPA charges is less than $0.05\,\theelcharge{}$, while it is about $0.2\,\theelcharge{}$ between the BPA and the MPA charges. The BPA and DPA charges thus show the same reasonable behavior with increasing electronegativity differences. 

\begin{figure}
\includegraphics[width=0.48\textwidth]{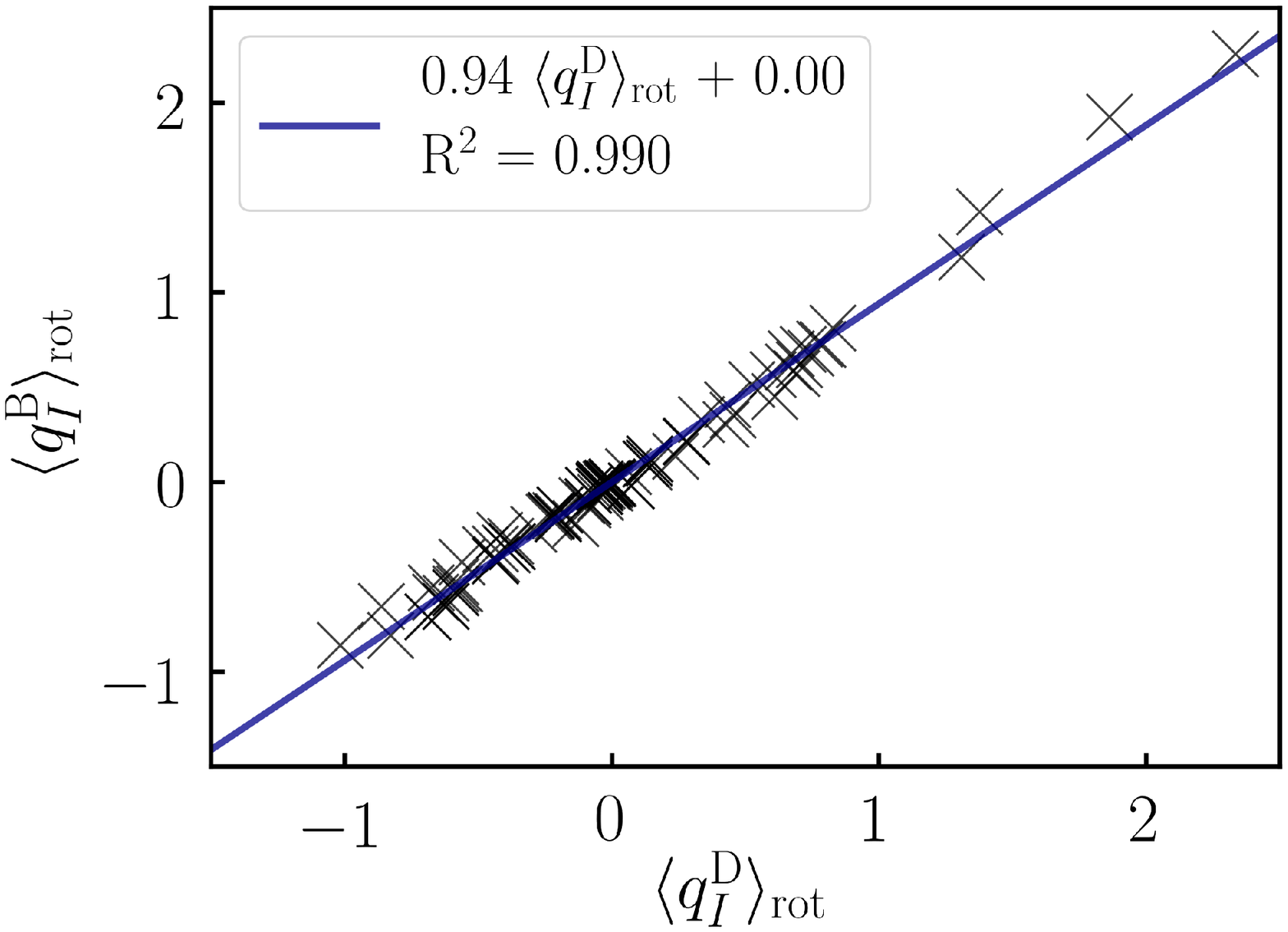} 
\caption{Linear regression between all rotationally averaged Berry (B) and atomic-polar-tensor (D) atomic charges ($\smash{q_I^\mathrm{B/D}}$, in $\theelcharge{}$) calculated in this work. The individual values are listed in the Supporting Information.}
\label{fig_corr}
\end{figure}

The absolute values of the BPA charges are slightly smaller than those of the DPA charges -- see Fig.~\ref{fig_corr}, where the slope of the linear regression between the two charges is 0.94. One possible reason for this difference is that the DPA charges include charge-transfer and/or polarization contributions, which are only partially included in the BPA charges, if at all.\cite{Richter2021} This explanation is supported by the observation that the Berry charge fluctuations and the total dipole moment tend to be larger when the BPA and the DPA charges differ more strongly. 

\begin{figure}
\includegraphics[width=0.48\textwidth]{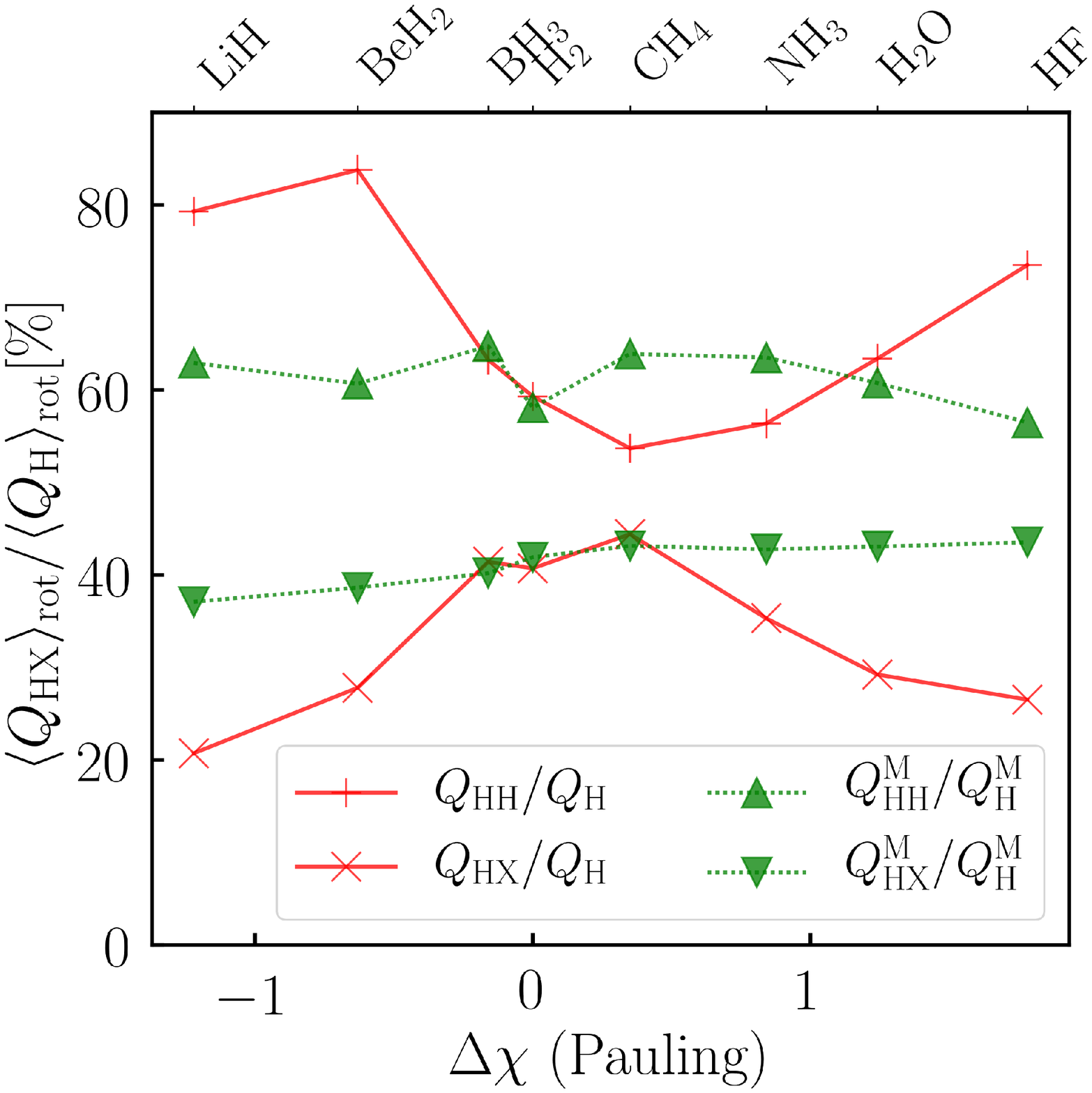} 
\caption{Comparison of rotationally averaged Berry charges ($Q_\mathrm{HX}$) and Mulliken overlap populations [$Q_\mathrm{HX}^\mathrm{M}$, see eq.~\eqref{bpa_001b}] for a series of H-containing molecules XH$_n$ (X = H, Li-F). For comparison, we divide both by the corresponding charges ($Q_\mathrm{H} = q_\mathrm{H}^\mathrm{B}-Z_\mathrm{H}$; $Q_\mathrm{H}^\mathrm{M} = q_\mathrm{H}^\mathrm{M}-Z_\mathrm{H}$) of hydrogen. The molecules are sorted according to the Pauling electronegativity difference ($\Delta \chi$) between X and H. When a molecule consists of more than one hydrogen atom, we display averaged values.}
\label{fig_op}
\end{figure}

Finally, we compare the rotationally averaged Berry charges with the Mulliken overlap populations. In Fig.~\ref{fig_op}, we plot these quantities divided by the atomic charges (without $\thez{\mathrm{H}}$) for a series of H-containing molecules of type XH$_n$. In this way, we obtain a measure for how much of the atomic charge on hydrogen stems from the atom itself [$\theq{\mathrm{H}}{\mathrm{H}}{\ther{}}/\theq{\mathrm{H}}{}{\ther{}}$] and from the neighboring atom X  [$\theq{\mathrm{H}}{\mathrm{X}}{\ther{}}/\theq{\mathrm{H}}{}{\ther{}}$] in the BPA and MPA schemes. A similar quantity cannot be obtained from the DPA without further assumptions. 

The BPA results agree with chemical intuition. For the ionic molecules LiH and FH, more than 70\% of the atomic charge of hydrogen can be assigned to hydrogen itself. This value decreases significantly as the HX bond becomes more covalent, reaching a minimum at 53\% for CH$_4$. The latter value correlates with the idea that two atoms \enquote{share} electrons in a covalent bond. This trend is not observed for the MPA, where the values are almost the same (about $60\%$) for all molecules. The MPA thus allows bonds to be identified but cannot distinguish between covalent and ionic bonds, unlike the BPA.

\section{Conclusions and Outlook}

In this work, we have shown that the Berry curvature can be rewritten in terms of the magnetic field and charge-like contributions as its role as the screening of the nuclei by the electrons implies. The resulting Berry charges and charge fluctuations reproduce about $90\%$ (and more than $95\%$ in some cases) of the exact Berry curvature, making them a good starting point for further approximations that aim at reducing the computational cost of \textit{ab initio} molecular dynamics in a strong magnetic field. Additionally, these charges can be used to construct atomic charges and overlap populations. The first results of this Berry population analysis (BPA) were encouraging, since the atomic charges were physically reasonable and close to the results of the widely used and well-established atomic-polar-tensor (DPA) charges. In addition, the overlap populations, which are not accessible in the DPA, give insight into the binding modes (covalent/ionic), while the Berry charge fluctuations indicate cases where a simple interpretation of the electronic structure in terms of atomic charges breaks down. We conclude that the BPA is a promising tool for the investigation of molecules, from which we expect to gain new insight into chemical bonds and electronic structure with and without a magnetic field.
\newpage
\newpage
\appendix

\section{Eigenvalues of the General $\Lambda$-Matrix}

We begin by introducing a coordinate system depending on the normalized interatomic distance vector $\therbar{\indnuci}{\indnucii}$ and the magnetic field $\theb{}$:
\begin{align}
\mathbf{e} &=
\begin{pmatrix} \therbar{\indnuci}{\indnucii} & \dfrac{\theb{}\times\therbar{\indnuci}{\indnucii}}{|\theb{}\times \therbar{\indnuci}{\indnucii}|} & \dfrac{\therbar{\indnuci}{\indnucii}\times[\theb{}\times\therbar{\indnuci}{\indnucii}]}{|\therbar{\indnuci}{\indnucii}\times[\theb{}\times \therbar{\indnuci}{\indnucii}]|}\end{pmatrix} 
\end{align}
Dropping the explicit dependence on $\ther{}$, we rewrite the matrices $\thebt{}$ and $\boldsymbol{\Gamma}_{IJ}^+$ in terms of these vectors using the angle $\theta$ between $\therbar{\indnuci}{\indnucii}$ and $\theb{}$
\begin{align}
\thebt{} &= 
|\theb{}|
\mathbf{e} 
\begin{pmatrix}
0 & - \sin(\theta) & 0 \\
\sin(\theta) & 0 & -\cos(\theta) \\
0 & \cos(\theta) & 0
\end{pmatrix} 
\mathbf{e}^\mathrm{T}  , \\
\boldsymbol{\Gamma}_{IJ}^+  &= 
|\theb{}|
\mathbf{e} 
\begin{pmatrix}
0 & \sin(\theta) & 0 \\
\sin(\theta) & 0 & 0\\
0 & 0 & 0
\end{pmatrix}
\mathbf{e}^\mathrm{T}  ,
\end{align}
so that $\boldsymbol{\Omega}_{IJ}^\mathrm{B2}$ takes the following form:
\begin{align}
&\boldsymbol{\Omega}_{IJ}^\mathrm{B2}
= - Q_{IJ} \thebt{}
- P_{IJ} \boldsymbol{\Gamma}_{IJ}^+ \nonumber \\
&=
|\theb{}|
\mathbf{e} 
\left(
\begin{smallmatrix}
0 & [Q_{IJ} - P_{IJ}]\sin(\theta) & 0 \\
-[Q_{IJ} + P_{IJ}]\sin(\theta) & 0 & Q_{IJ} \cos(\theta) \\
0 & -Q_{IJ} \cos(\theta) & 0
\end{smallmatrix} 
\right)
\mathbf{e}^\mathrm{T} .
\end{align}
From this we can now obtain the eigenvalues of \begin{equation}
\boldsymbol{\Lambda}_{IJ}= |\theb{}|^{-2} \left[ \boldsymbol{\Omega}_{IJ}^\mathrm{B2}\right]^\mathrm{T}
\boldsymbol{\Omega}_{IJ}^\mathrm{B2}
\end{equation}
as:
\begin{align}
\lambda_{\indnuci\indnucii}^1  &= 0,\\
\lambda_{\indnuci\indnucii}^{2/3}  &= Q_{IJ}^2
+ \sin^2(\theta)P_{IJ}^2
\pm 
2 Q_{IJ} P_{IJ}
\sin^2(\theta).
\end{align}

\section*{Supplementary Material}

See the supplementary material for the convergence study of the numerical spherical integration and the list of all atomic charges plotted in this work.

\section*{Conflict of Interest}

The authors declare no competing financial interest.

\section*{Acknowledgments}

This work was supported by the Research Council of Norway through ‘‘Magnetic Chemistry’’ Grant No.\,287950 and CoE Hylleraas Centre for Quantum Molecular Sciences Grant No.\,262695. This work has also received support from the Norwegian Supercomputing Program (NOTUR) through a grant of computer time (Grant No.\ NN4654K).

\clearpage

\begin{thebibliography}{65}%
\makeatletter
\providecommand \@ifxundefined [1]{%
 \@ifx{#1\undefined}
}%
\providecommand \@ifnum [1]{%
 \ifnum #1\expandafter \@firstoftwo
 \else \expandafter \@secondoftwo
 \fi
}%
\providecommand \@ifx [1]{%
 \ifx #1\expandafter \@firstoftwo
 \else \expandafter \@secondoftwo
 \fi
}%
\providecommand \natexlab [1]{#1}%
\providecommand \enquote  [1]{``#1''}%
\providecommand \bibnamefont  [1]{#1}%
\providecommand \bibfnamefont [1]{#1}%
\providecommand \citenamefont [1]{#1}%
\providecommand \href@noop [0]{\@secondoftwo}%
\providecommand \href [0]{\begingroup \@sanitize@url \@href}%
\providecommand \@href[1]{\@@startlink{#1}\@@href}%
\providecommand \@@href[1]{\endgroup#1\@@endlink}%
\providecommand \@sanitize@url [0]{\catcode `\\12\catcode `\$12\catcode
  `\&12\catcode `\#12\catcode `\^12\catcode `\_12\catcode `\%12\relax}%
\providecommand \@@startlink[1]{}%
\providecommand \@@endlink[0]{}%
\providecommand \url  [0]{\begingroup\@sanitize@url \@url }%
\providecommand \@url [1]{\endgroup\@href {#1}{\urlprefix }}%
\providecommand \urlprefix  [0]{URL }%
\providecommand \Eprint [0]{\href }%
\providecommand \doibase [0]{http://dx.doi.org/}%
\providecommand \selectlanguage [0]{\@gobble}%
\providecommand \bibinfo  [0]{\@secondoftwo}%
\providecommand \bibfield  [0]{\@secondoftwo}%
\providecommand \translation [1]{[#1]}%
\providecommand \BibitemOpen [0]{}%
\providecommand \bibitemStop [0]{}%
\providecommand \bibitemNoStop [0]{.\EOS\space}%
\providecommand \EOS [0]{\spacefactor3000\relax}%
\providecommand \BibitemShut  [1]{\csname bibitem#1\endcsname}%
\let\auto@bib@innerbib\@empty
\bibitem [{\citenamefont {Wiberg}\ and\ \citenamefont
  {Rablen}(1993)}]{Wiberg1993}%
  \BibitemOpen
  \bibfield  {author} {\bibinfo {author} {\bibfnamefont {K.~B.}\ \bibnamefont
  {Wiberg}}\ and\ \bibinfo {author} {\bibfnamefont {P.~R.}\ \bibnamefont
  {Rablen}},\ }\href@noop {} {\bibfield  {journal} {\bibinfo  {journal} {J.
  Comput. Chem.}\ }\textbf {\bibinfo {volume} {14}},\ \bibinfo {pages} {1504}
  (\bibinfo {year} {1993})}\BibitemShut {NoStop}%
\bibitem [{\citenamefont {Meister}\ and\ \citenamefont
  {Schwarz}(1994)}]{Meister1994}%
  \BibitemOpen
  \bibfield  {author} {\bibinfo {author} {\bibfnamefont {J.}~\bibnamefont
  {Meister}}\ and\ \bibinfo {author} {\bibfnamefont {W.~H.~E.}\ \bibnamefont
  {Schwarz}},\ }\href@noop {} {\bibfield  {journal} {\bibinfo  {journal} {J.
  Chem. Phys}\ }\textbf {\bibinfo {volume} {98}},\ \bibinfo {pages} {8245}
  (\bibinfo {year} {1994})}\BibitemShut {NoStop}%
\bibitem [{\citenamefont {Cramer}(2004)}]{Cramer2004}%
  \BibitemOpen
  \bibfield  {author} {\bibinfo {author} {\bibfnamefont {C.~J.}\ \bibnamefont
  {Cramer}},\ }\href@noop {} {\emph {\bibinfo {title} {{Essentials of
  Computational Chemistry: Theories and Models}}}}\ (\bibinfo  {publisher}
  {Wiley},\ \bibinfo {year} {2004})\BibitemShut {NoStop}%
\bibitem [{\citenamefont {Cho}\ \emph {et~al.}(2020)\citenamefont {Cho},
  \citenamefont {Sylvetsky}, \citenamefont {Eshafi}, \citenamefont {Santra},
  \citenamefont {Efremenko},\ and\ \citenamefont {Martin}}]{Cho2020}%
  \BibitemOpen
  \bibfield  {author} {\bibinfo {author} {\bibfnamefont {M.}~\bibnamefont
  {Cho}}, \bibinfo {author} {\bibfnamefont {N.}~\bibnamefont {Sylvetsky}},
  \bibinfo {author} {\bibfnamefont {S.}~\bibnamefont {Eshafi}}, \bibinfo
  {author} {\bibfnamefont {G.}~\bibnamefont {Santra}}, \bibinfo {author}
  {\bibfnamefont {I.}~\bibnamefont {Efremenko}}, \ and\ \bibinfo {author}
  {\bibfnamefont {J.~M.}\ \bibnamefont {Martin}},\ }\href@noop {} {\bibfield
  {journal} {\bibinfo  {journal} {ChemPhysChem}\ }\textbf {\bibinfo {volume}
  {21}},\ \bibinfo {pages} {688} (\bibinfo {year} {2020})}\BibitemShut
  {NoStop}%
\bibitem [{\citenamefont {Mulliken}(1955{\natexlab{a}})}]{Mulliken1955}%
  \BibitemOpen
  \bibfield  {author} {\bibinfo {author} {\bibfnamefont {R.~S.}\ \bibnamefont
  {Mulliken}},\ }\href@noop {} {\bibfield  {journal} {\bibinfo  {journal} {J.
  Chem. Phys.}\ }\textbf {\bibinfo {volume} {23}},\ \bibinfo {pages} {1833}
  (\bibinfo {year} {1955}{\natexlab{a}})}\BibitemShut {NoStop}%
\bibitem [{\citenamefont {Mulliken}(1955{\natexlab{b}})}]{Mulliken1955a}%
  \BibitemOpen
  \bibfield  {author} {\bibinfo {author} {\bibfnamefont {R.~S.}\ \bibnamefont
  {Mulliken}},\ }\href@noop {} {\bibfield  {journal} {\bibinfo  {journal} {J.
  Chem. Phys.}\ }\textbf {\bibinfo {volume} {23}},\ \bibinfo {pages} {1841}
  (\bibinfo {year} {1955}{\natexlab{b}})}\BibitemShut {NoStop}%
\bibitem [{\citenamefont {Mulliken}(1955{\natexlab{c}})}]{Mulliken1955b}%
  \BibitemOpen
  \bibfield  {author} {\bibinfo {author} {\bibfnamefont {R.~S.}\ \bibnamefont
  {Mulliken}},\ }\href@noop {} {\bibfield  {journal} {\bibinfo  {journal} {J.
  Chem. Phys.}\ }\textbf {\bibinfo {volume} {23}},\ \bibinfo {pages} {2338}
  (\bibinfo {year} {1955}{\natexlab{c}})}\BibitemShut {NoStop}%
\bibitem [{\citenamefont {L{\"{o}}wdin}(1950)}]{Lowdin1950}%
  \BibitemOpen
  \bibfield  {author} {\bibinfo {author} {\bibfnamefont {P.~O.}\ \bibnamefont
  {L{\"{o}}wdin}},\ }\href@noop {} {\bibfield  {journal} {\bibinfo  {journal}
  {J. Chem. Phys.}\ }\textbf {\bibinfo {volume} {18}},\ \bibinfo {pages} {365}
  (\bibinfo {year} {1950})}\BibitemShut {NoStop}%
\bibitem [{\citenamefont {Baker}(1985)}]{Baker1985}%
  \BibitemOpen
  \bibfield  {author} {\bibinfo {author} {\bibfnamefont {J.}~\bibnamefont
  {Baker}},\ }\href@noop {} {\bibfield  {journal} {\bibinfo  {journal} {Theor.
  Chim. Acta}\ }\textbf {\bibinfo {volume} {68}},\ \bibinfo {pages} {221}
  (\bibinfo {year} {1985})}\BibitemShut {NoStop}%
\bibitem [{\citenamefont {Reed}, \citenamefont {Weinstock},\ and\ \citenamefont
  {Weinhold}(1985)}]{Reed1985}%
  \BibitemOpen
  \bibfield  {author} {\bibinfo {author} {\bibfnamefont {A.~E.}\ \bibnamefont
  {Reed}}, \bibinfo {author} {\bibfnamefont {R.~B.}\ \bibnamefont {Weinstock}},
  \ and\ \bibinfo {author} {\bibfnamefont {F.}~\bibnamefont {Weinhold}},\
  }\href@noop {} {\bibfield  {journal} {\bibinfo  {journal} {J. Chem. Phys.}\
  }\textbf {\bibinfo {volume} {83}},\ \bibinfo {pages} {735} (\bibinfo {year}
  {1985})}\BibitemShut {NoStop}%
\bibitem [{\citenamefont {Maslen}\ and\ \citenamefont
  {Spackman}(1985)}]{Maslen1985}%
  \BibitemOpen
  \bibfield  {author} {\bibinfo {author} {\bibfnamefont {E.}~\bibnamefont
  {Maslen}}\ and\ \bibinfo {author} {\bibfnamefont {M.}~\bibnamefont
  {Spackman}},\ }\href@noop {} {\bibfield  {journal} {\bibinfo  {journal}
  {Aust. J. Phys.}\ }\textbf {\bibinfo {volume} {38}},\ \bibinfo {pages} {273}
  (\bibinfo {year} {1985})}\BibitemShut {NoStop}%
\bibitem [{\citenamefont {Hirshfeld}(1977)}]{Hirshfeld1977}%
  \BibitemOpen
  \bibfield  {author} {\bibinfo {author} {\bibfnamefont {F.~L.}\ \bibnamefont
  {Hirshfeld}},\ }\href@noop {} {\bibfield  {journal} {\bibinfo  {journal}
  {Theor. Chim. Acta}\ }\textbf {\bibinfo {volume} {44}},\ \bibinfo {pages}
  {129} (\bibinfo {year} {1977})}\BibitemShut {NoStop}%
\bibitem [{\citenamefont {Marenich}\ \emph {et~al.}(2012)\citenamefont
  {Marenich}, \citenamefont {Jerome}, \citenamefont {Cramer},\ and\
  \citenamefont {Truhlar}}]{Marenich2012}%
  \BibitemOpen
  \bibfield  {author} {\bibinfo {author} {\bibfnamefont {A.~V.}\ \bibnamefont
  {Marenich}}, \bibinfo {author} {\bibfnamefont {S.~V.}\ \bibnamefont
  {Jerome}}, \bibinfo {author} {\bibfnamefont {C.~J.}\ \bibnamefont {Cramer}},
  \ and\ \bibinfo {author} {\bibfnamefont {D.~G.}\ \bibnamefont {Truhlar}},\
  }\href@noop {} {\bibfield  {journal} {\bibinfo  {journal} {J. Chem. Theory
  Comput.}\ }\textbf {\bibinfo {volume} {8}},\ \bibinfo {pages} {527} (\bibinfo
  {year} {2012})}\BibitemShut {NoStop}%
\bibitem [{\citenamefont {Manz}\ and\ \citenamefont {Limas}(2016)}]{Manz2016}%
  \BibitemOpen
  \bibfield  {author} {\bibinfo {author} {\bibfnamefont {T.~A.}\ \bibnamefont
  {Manz}}\ and\ \bibinfo {author} {\bibfnamefont {N.~G.}\ \bibnamefont
  {Limas}},\ }\href@noop {} {\bibfield  {journal} {\bibinfo  {journal} {RSC
  Adv.}\ }\textbf {\bibinfo {volume} {6}},\ \bibinfo {pages} {47771} (\bibinfo
  {year} {2016})}\BibitemShut {NoStop}%
\bibitem [{\citenamefont {Limas}\ and\ \citenamefont {Manz}(2016)}]{Limas2016}%
  \BibitemOpen
  \bibfield  {author} {\bibinfo {author} {\bibfnamefont {N.~G.}\ \bibnamefont
  {Limas}}\ and\ \bibinfo {author} {\bibfnamefont {T.~A.}\ \bibnamefont
  {Manz}},\ }\href@noop {} {\bibfield  {journal} {\bibinfo  {journal} {RSC
  Adv.}\ }\textbf {\bibinfo {volume} {6}},\ \bibinfo {pages} {45727} (\bibinfo
  {year} {2016})}\BibitemShut {NoStop}%
\bibitem [{\citenamefont {Bayly}\ \emph {et~al.}(1993)\citenamefont {Bayly},
  \citenamefont {Cieplak}, \citenamefont {Cornell},\ and\ \citenamefont
  {Kollman}}]{Bayly1993}%
  \BibitemOpen
  \bibfield  {author} {\bibinfo {author} {\bibfnamefont {C.~I.}\ \bibnamefont
  {Bayly}}, \bibinfo {author} {\bibfnamefont {P.}~\bibnamefont {Cieplak}},
  \bibinfo {author} {\bibfnamefont {W.~D.}\ \bibnamefont {Cornell}}, \ and\
  \bibinfo {author} {\bibfnamefont {P.~A.}\ \bibnamefont {Kollman}},\
  }\href@noop {} {\bibfield  {journal} {\bibinfo  {journal} {J. Phys. Chem.}\
  }\textbf {\bibinfo {volume} {97}},\ \bibinfo {pages} {10269} (\bibinfo {year}
  {1993})}\BibitemShut {NoStop}%
\bibitem [{\citenamefont {Cioslowski}(1989)}]{Cioslowski1989}%
  \BibitemOpen
  \bibfield  {author} {\bibinfo {author} {\bibfnamefont {J.}~\bibnamefont
  {Cioslowski}},\ }\href@noop {} {\bibfield  {journal} {\bibinfo  {journal} {J.
  Am. Chem. Soc.}\ }\textbf {\bibinfo {volume} {111}},\ \bibinfo {pages} {8333}
  (\bibinfo {year} {1989})}\BibitemShut {NoStop}%
\bibitem [{\citenamefont {Cioslowski}\ \emph {et~al.}(1990)\citenamefont
  {Cioslowski}, \citenamefont {Hamilton}, \citenamefont {Gustavo},
  \citenamefont {{Andes Hess}}, \citenamefont {Hu}, \citenamefont {Schaad},\
  and\ \citenamefont {Dupuis}}]{Cioslowski1990}%
  \BibitemOpen
  \bibfield  {author} {\bibinfo {author} {\bibfnamefont {J.}~\bibnamefont
  {Cioslowski}}, \bibinfo {author} {\bibfnamefont {T.}~\bibnamefont
  {Hamilton}}, \bibinfo {author} {\bibfnamefont {S.}~\bibnamefont {Gustavo}},
  \bibinfo {author} {\bibfnamefont {B.}~\bibnamefont {{Andes Hess}}}, \bibinfo
  {author} {\bibfnamefont {J.}~\bibnamefont {Hu}}, \bibinfo {author}
  {\bibfnamefont {L.~J.}\ \bibnamefont {Schaad}}, \ and\ \bibinfo {author}
  {\bibfnamefont {M.}~\bibnamefont {Dupuis}},\ }\href@noop {} {\bibfield
  {journal} {\bibinfo  {journal} {J. Am. Chem. Soc.}\ }\textbf {\bibinfo
  {volume} {112}},\ \bibinfo {pages} {4183} (\bibinfo {year}
  {1990})}\BibitemShut {NoStop}%
\bibitem [{\citenamefont {Haaland}\ \emph {et~al.}(2000)\citenamefont
  {Haaland}, \citenamefont {Helgaker}, \citenamefont {Ruud},\ and\
  \citenamefont {Shorokhov}}]{Haaland2000}%
  \BibitemOpen
  \bibfield  {author} {\bibinfo {author} {\bibfnamefont {A.}~\bibnamefont
  {Haaland}}, \bibinfo {author} {\bibfnamefont {T.}~\bibnamefont {Helgaker}},
  \bibinfo {author} {\bibfnamefont {K.}~\bibnamefont {Ruud}}, \ and\ \bibinfo
  {author} {\bibfnamefont {D.~J.}\ \bibnamefont {Shorokhov}},\ }\href@noop {}
  {\bibfield  {journal} {\bibinfo  {journal} {Res. Sci. Educ.}\ }\textbf
  {\bibinfo {volume} {77}},\ \bibinfo {pages} {1076} (\bibinfo {year}
  {2000})}\BibitemShut {NoStop}%
\bibitem [{\citenamefont {Shukla}(2000)}]{Shukla2000}%
  \BibitemOpen
  \bibfield  {author} {\bibinfo {author} {\bibfnamefont {A.}~\bibnamefont
  {Shukla}},\ }\href@noop {} {\bibfield  {journal} {\bibinfo  {journal} {Phys.
  Rev. B - Condens. Matter Mater. Phys.}\ }\textbf {\bibinfo {volume} {61}},\
  \bibinfo {pages} {13277} (\bibinfo {year} {2000})}\BibitemShut {NoStop}%
\bibitem [{\citenamefont {Milani}\ and\ \citenamefont
  {Castiglioni}(2010)}]{Milani2010}%
  \BibitemOpen
  \bibfield  {author} {\bibinfo {author} {\bibfnamefont {A.}~\bibnamefont
  {Milani}}\ and\ \bibinfo {author} {\bibfnamefont {C.}~\bibnamefont
  {Castiglioni}},\ }\href@noop {} {\bibfield  {journal} {\bibinfo  {journal}
  {J. Mol. Struct. THEOCHEM}\ }\textbf {\bibinfo {volume} {955}},\ \bibinfo
  {pages} {158} (\bibinfo {year} {2010})}\BibitemShut {NoStop}%
\bibitem [{\citenamefont {Peters}\ \emph {et~al.}(2021)\citenamefont {Peters},
  \citenamefont {Culpitt}, \citenamefont {Monzel}, \citenamefont {Tellgren},\
  and\ \citenamefont {Helgaker}}]{Peters2021}%
  \BibitemOpen
  \bibfield  {author} {\bibinfo {author} {\bibfnamefont {L.~D.~M.}\
  \bibnamefont {Peters}}, \bibinfo {author} {\bibfnamefont {T.}~\bibnamefont
  {Culpitt}}, \bibinfo {author} {\bibfnamefont {L.}~\bibnamefont {Monzel}},
  \bibinfo {author} {\bibfnamefont {E.~I.}\ \bibnamefont {Tellgren}}, \ and\
  \bibinfo {author} {\bibfnamefont {T.}~\bibnamefont {Helgaker}},\ }\href@noop
  {} {\bibfield  {journal} {\bibinfo  {journal} {J. Chem. Phys.}\ }\textbf
  {\bibinfo {volume} {155}},\ \bibinfo {pages} {024105} (\bibinfo {year}
  {2021})}\BibitemShut {NoStop}%
\bibitem [{\citenamefont {Monzel}\ \emph {et~al.}(2022)\citenamefont {Monzel},
  \citenamefont {Pausch}, \citenamefont {Peters}, \citenamefont {Tellgren},
  \citenamefont {Helgaker},\ and\ \citenamefont {Klopper}}]{Monzel2022}%
  \BibitemOpen
  \bibfield  {author} {\bibinfo {author} {\bibfnamefont {L.}~\bibnamefont
  {Monzel}}, \bibinfo {author} {\bibfnamefont {A.}~\bibnamefont {Pausch}},
  \bibinfo {author} {\bibfnamefont {L.~D.~M.}\ \bibnamefont {Peters}}, \bibinfo
  {author} {\bibfnamefont {E.~I.}\ \bibnamefont {Tellgren}}, \bibinfo {author}
  {\bibfnamefont {T.}~\bibnamefont {Helgaker}}, \ and\ \bibinfo {author}
  {\bibfnamefont {W.}~\bibnamefont {Klopper}},\ }\href@noop {} {\bibfield
  {journal} {\bibinfo  {journal} {J. Chem. Phys.}\ }\textbf {\bibinfo {volume}
  {157}},\ \bibinfo {pages} {054106} (\bibinfo {year} {2022})}\BibitemShut
  {NoStop}%
\bibitem [{\citenamefont {Tellgren}, \citenamefont {Soncini},\ and\
  \citenamefont {Helgaker}(2008)}]{Tellgren2008}%
  \BibitemOpen
  \bibfield  {author} {\bibinfo {author} {\bibfnamefont {E.~I.}\ \bibnamefont
  {Tellgren}}, \bibinfo {author} {\bibfnamefont {A.}~\bibnamefont {Soncini}}, \
  and\ \bibinfo {author} {\bibfnamefont {T.}~\bibnamefont {Helgaker}},\
  }\href@noop {} {\bibfield  {journal} {\bibinfo  {journal} {J. Chem. Phys.}\
  }\textbf {\bibinfo {volume} {129}},\ \bibinfo {pages} {154114} (\bibinfo
  {year} {2008})}\BibitemShut {NoStop}%
\bibitem [{\citenamefont {Tellgren}, \citenamefont {Helgaker},\ and\
  \citenamefont {Soncini}(2009)}]{Tellgren2009}%
  \BibitemOpen
  \bibfield  {author} {\bibinfo {author} {\bibfnamefont {E.~I.}\ \bibnamefont
  {Tellgren}}, \bibinfo {author} {\bibfnamefont {T.}~\bibnamefont {Helgaker}},
  \ and\ \bibinfo {author} {\bibfnamefont {A.}~\bibnamefont {Soncini}},\
  }\href@noop {} {\bibfield  {journal} {\bibinfo  {journal} {Phys. Chem. Chem.
  Phys.}\ }\textbf {\bibinfo {volume} {11}},\ \bibinfo {pages} {5489} (\bibinfo
  {year} {2009})}\BibitemShut {NoStop}%
\bibitem [{\citenamefont {Lange}\ \emph {et~al.}(2012)\citenamefont {Lange},
  \citenamefont {Tellgren}, \citenamefont {Hoffmann},\ and\ \citenamefont
  {Helgaker}}]{Lange2012}%
  \BibitemOpen
  \bibfield  {author} {\bibinfo {author} {\bibfnamefont {K.~K.}\ \bibnamefont
  {Lange}}, \bibinfo {author} {\bibfnamefont {E.~I.}\ \bibnamefont {Tellgren}},
  \bibinfo {author} {\bibfnamefont {M.~R.}\ \bibnamefont {Hoffmann}}, \ and\
  \bibinfo {author} {\bibfnamefont {T.}~\bibnamefont {Helgaker}},\ }\href@noop
  {} {\bibfield  {journal} {\bibinfo  {journal} {Science}\ }\textbf {\bibinfo
  {volume} {337}},\ \bibinfo {pages} {327} (\bibinfo {year}
  {2012})}\BibitemShut {NoStop}%
\bibitem [{\citenamefont {Tellgren}, \citenamefont {Reine},\ and\ \citenamefont
  {Helgaker}(2012)}]{Tellgren2012}%
  \BibitemOpen
  \bibfield  {author} {\bibinfo {author} {\bibfnamefont {E.~I.}\ \bibnamefont
  {Tellgren}}, \bibinfo {author} {\bibfnamefont {S.~S.}\ \bibnamefont {Reine}},
  \ and\ \bibinfo {author} {\bibfnamefont {T.}~\bibnamefont {Helgaker}},\
  }\href@noop {} {\bibfield  {journal} {\bibinfo  {journal} {Phys. Chem. Chem.
  Phys.}\ }\textbf {\bibinfo {volume} {14}},\ \bibinfo {pages} {9492} (\bibinfo
  {year} {2012})}\BibitemShut {NoStop}%
\bibitem [{\citenamefont {Reynolds}\ and\ \citenamefont
  {Shiozaki}(2015)}]{Reynolds2015}%
  \BibitemOpen
  \bibfield  {author} {\bibinfo {author} {\bibfnamefont {R.~D.}\ \bibnamefont
  {Reynolds}}\ and\ \bibinfo {author} {\bibfnamefont {T.}~\bibnamefont
  {Shiozaki}},\ }\href@noop {} {\bibfield  {journal} {\bibinfo  {journal}
  {Phys. Chem. Chem. Phys.}\ }\textbf {\bibinfo {volume} {17}},\ \bibinfo
  {pages} {14280} (\bibinfo {year} {2015})}\BibitemShut {NoStop}%
\bibitem [{\citenamefont {Stopkowicz}\ \emph {et~al.}(2015)\citenamefont
  {Stopkowicz}, \citenamefont {Gauss}, \citenamefont {Lange}, \citenamefont
  {Tellgren},\ and\ \citenamefont {Helgaker}}]{Stopkowicz2015}%
  \BibitemOpen
  \bibfield  {author} {\bibinfo {author} {\bibfnamefont {S.}~\bibnamefont
  {Stopkowicz}}, \bibinfo {author} {\bibfnamefont {J.}~\bibnamefont {Gauss}},
  \bibinfo {author} {\bibfnamefont {K.~K.}\ \bibnamefont {Lange}}, \bibinfo
  {author} {\bibfnamefont {E.~I.}\ \bibnamefont {Tellgren}}, \ and\ \bibinfo
  {author} {\bibfnamefont {T.}~\bibnamefont {Helgaker}},\ }\href@noop {}
  {\bibfield  {journal} {\bibinfo  {journal} {J. Chem. Phys.}\ }\textbf
  {\bibinfo {volume} {143}},\ \bibinfo {pages} {074110} (\bibinfo {year}
  {2015})}\BibitemShut {NoStop}%
\bibitem [{\citenamefont {Hampe}\ and\ \citenamefont
  {Stopkowicz}(2017)}]{Hampe2017}%
  \BibitemOpen
  \bibfield  {author} {\bibinfo {author} {\bibfnamefont {F.}~\bibnamefont
  {Hampe}}\ and\ \bibinfo {author} {\bibfnamefont {S.}~\bibnamefont
  {Stopkowicz}},\ }\href@noop {} {\bibfield  {journal} {\bibinfo  {journal} {J.
  Chem. Phys.}\ }\textbf {\bibinfo {volume} {146}},\ \bibinfo {pages} {154105}
  (\bibinfo {year} {2017})}\BibitemShut {NoStop}%
\bibitem [{\citenamefont {Irons}, \citenamefont {Zemen},\ and\ \citenamefont
  {Teale}(2017)}]{Irons2017}%
  \BibitemOpen
  \bibfield  {author} {\bibinfo {author} {\bibfnamefont {T.~J.~P.}\
  \bibnamefont {Irons}}, \bibinfo {author} {\bibfnamefont {J.}~\bibnamefont
  {Zemen}}, \ and\ \bibinfo {author} {\bibfnamefont {A.~M.}\ \bibnamefont
  {Teale}},\ }\href@noop {} {\bibfield  {journal} {\bibinfo  {journal} {J.
  Chem. Theory Comput.}\ }\textbf {\bibinfo {volume} {13}},\ \bibinfo {pages}
  {3636} (\bibinfo {year} {2017})}\BibitemShut {NoStop}%
\bibitem [{\citenamefont {Hampe}\ and\ \citenamefont
  {Stopkowicz}(2019)}]{Hampe2019}%
  \BibitemOpen
  \bibfield  {author} {\bibinfo {author} {\bibfnamefont {F.}~\bibnamefont
  {Hampe}}\ and\ \bibinfo {author} {\bibfnamefont {S.}~\bibnamefont
  {Stopkowicz}},\ }\href@noop {} {\bibfield  {journal} {\bibinfo  {journal} {J.
  Chem. Theory Comput.}\ }\textbf {\bibinfo {volume} {15}},\ \bibinfo {pages}
  {4036} (\bibinfo {year} {2019})}\BibitemShut {NoStop}%
\bibitem [{\citenamefont {Sen}, \citenamefont {Lange},\ and\ \citenamefont
  {Tellgren}(2019)}]{Sen2019}%
  \BibitemOpen
  \bibfield  {author} {\bibinfo {author} {\bibfnamefont {S.}~\bibnamefont
  {Sen}}, \bibinfo {author} {\bibfnamefont {K.~K.}\ \bibnamefont {Lange}}, \
  and\ \bibinfo {author} {\bibfnamefont {E.~I.}\ \bibnamefont {Tellgren}},\
  }\href@noop {} {\bibfield  {journal} {\bibinfo  {journal} {J. Chem. Theory
  Comput.}\ }\textbf {\bibinfo {volume} {15}},\ \bibinfo {pages} {3974}
  (\bibinfo {year} {2019})}\BibitemShut {NoStop}%
\bibitem [{\citenamefont {Sun}\ \emph {et~al.}(2019)\citenamefont {Sun},
  \citenamefont {Williams-Young}, \citenamefont {Stetina},\ and\ \citenamefont
  {Li}}]{Sun2019}%
  \BibitemOpen
  \bibfield  {author} {\bibinfo {author} {\bibfnamefont {S.}~\bibnamefont
  {Sun}}, \bibinfo {author} {\bibfnamefont {D.~B.}\ \bibnamefont
  {Williams-Young}}, \bibinfo {author} {\bibfnamefont {T.~F.}\ \bibnamefont
  {Stetina}}, \ and\ \bibinfo {author} {\bibfnamefont {X.}~\bibnamefont {Li}},\
  }\href@noop {} {\bibfield  {journal} {\bibinfo  {journal} {J. Chem. Theory
  Comput.}\ }\textbf {\bibinfo {volume} {15}},\ \bibinfo {pages} {348}
  (\bibinfo {year} {2019})}\BibitemShut {NoStop}%
\bibitem [{\citenamefont {Austad}\ \emph {et~al.}(2020)\citenamefont {Austad},
  \citenamefont {Borgoo}, \citenamefont {Tellgren},\ and\ \citenamefont
  {Helgaker}}]{Austad2020}%
  \BibitemOpen
  \bibfield  {author} {\bibinfo {author} {\bibfnamefont {J.}~\bibnamefont
  {Austad}}, \bibinfo {author} {\bibfnamefont {A.}~\bibnamefont {Borgoo}},
  \bibinfo {author} {\bibfnamefont {E.~I.}\ \bibnamefont {Tellgren}}, \ and\
  \bibinfo {author} {\bibfnamefont {T.}~\bibnamefont {Helgaker}},\ }\href@noop
  {} {\bibfield  {journal} {\bibinfo  {journal} {Phys. Chem. Chem. Phys.}\
  }\textbf {\bibinfo {volume} {22}},\ \bibinfo {pages} {23502} (\bibinfo {year}
  {2020})}\BibitemShut {NoStop}%
\bibitem [{\citenamefont {Hampe}, \citenamefont {Gross},\ and\ \citenamefont
  {Stopkowicz}(2020)}]{Hampe2020}%
  \BibitemOpen
  \bibfield  {author} {\bibinfo {author} {\bibfnamefont {F.}~\bibnamefont
  {Hampe}}, \bibinfo {author} {\bibfnamefont {N.}~\bibnamefont {Gross}}, \ and\
  \bibinfo {author} {\bibfnamefont {S.}~\bibnamefont {Stopkowicz}},\
  }\href@noop {} {\bibfield  {journal} {\bibinfo  {journal} {Phys. Chem. Chem.
  Phys.}\ }\textbf {\bibinfo {volume} {22}},\ \bibinfo {pages} {23522}
  (\bibinfo {year} {2020})}\BibitemShut {NoStop}%
\bibitem [{\citenamefont {Pausch}\ and\ \citenamefont
  {Klopper}(2020)}]{Pausch2020}%
  \BibitemOpen
  \bibfield  {author} {\bibinfo {author} {\bibfnamefont {A.}~\bibnamefont
  {Pausch}}\ and\ \bibinfo {author} {\bibfnamefont {W.}~\bibnamefont
  {Klopper}},\ }\href@noop {} {\bibfield  {journal} {\bibinfo  {journal} {Mol.
  Phys.}\ }\textbf {\bibinfo {volume} {118}},\ \bibinfo {pages} {e1736675}
  (\bibinfo {year} {2020})}\BibitemShut {NoStop}%
\bibitem [{\citenamefont {Williams-Young}\ \emph {et~al.}(2020)\citenamefont
  {Williams-Young}, \citenamefont {Petrone}, \citenamefont {Sun}, \citenamefont
  {Stetina}, \citenamefont {Lestrange}, \citenamefont {Hoyer}, \citenamefont
  {Nascimento}, \citenamefont {Koulias}, \citenamefont {Wildman}, \citenamefont
  {Kasper}, \citenamefont {Goings}, \citenamefont {Ding}, \citenamefont
  {DePrince}, \citenamefont {Valeev},\ and\ \citenamefont
  {Li}}]{Williams-Young2020}%
  \BibitemOpen
  \bibfield  {author} {\bibinfo {author} {\bibfnamefont {D.~B.}\ \bibnamefont
  {Williams-Young}}, \bibinfo {author} {\bibfnamefont {A.}~\bibnamefont
  {Petrone}}, \bibinfo {author} {\bibfnamefont {S.}~\bibnamefont {Sun}},
  \bibinfo {author} {\bibfnamefont {T.~F.}\ \bibnamefont {Stetina}}, \bibinfo
  {author} {\bibfnamefont {P.}~\bibnamefont {Lestrange}}, \bibinfo {author}
  {\bibfnamefont {C.~E.}\ \bibnamefont {Hoyer}}, \bibinfo {author}
  {\bibfnamefont {D.~R.}\ \bibnamefont {Nascimento}}, \bibinfo {author}
  {\bibfnamefont {L.}~\bibnamefont {Koulias}}, \bibinfo {author} {\bibfnamefont
  {A.}~\bibnamefont {Wildman}}, \bibinfo {author} {\bibfnamefont
  {J.}~\bibnamefont {Kasper}}, \bibinfo {author} {\bibfnamefont {J.~J.}\
  \bibnamefont {Goings}}, \bibinfo {author} {\bibfnamefont {F.}~\bibnamefont
  {Ding}}, \bibinfo {author} {\bibfnamefont {A.~E.}\ \bibnamefont {DePrince}},
  \bibinfo {author} {\bibfnamefont {E.~F.}\ \bibnamefont {Valeev}}, \ and\
  \bibinfo {author} {\bibfnamefont {X.}~\bibnamefont {Li}},\ }\href@noop {}
  {\bibfield  {journal} {\bibinfo  {journal} {Wiley Interdiscip. Rev. Comput.
  Mol. Sci.}\ }\textbf {\bibinfo {volume} {10}},\ \bibinfo {pages} {e1436}
  (\bibinfo {year} {2020})}\BibitemShut {NoStop}%
\bibitem [{\citenamefont {Irons}, \citenamefont {David},\ and\ \citenamefont
  {Teale}(2021)}]{Irons2021}%
  \BibitemOpen
  \bibfield  {author} {\bibinfo {author} {\bibfnamefont {T.~J.}\ \bibnamefont
  {Irons}}, \bibinfo {author} {\bibfnamefont {G.}~\bibnamefont {David}}, \ and\
  \bibinfo {author} {\bibfnamefont {A.~M.}\ \bibnamefont {Teale}},\ }\href@noop
  {} {\bibfield  {journal} {\bibinfo  {journal} {J. Chem. Theory Comput.}\
  }\textbf {\bibinfo {volume} {17}},\ \bibinfo {pages} {2166} (\bibinfo {year}
  {2021})}\BibitemShut {NoStop}%
\bibitem [{\citenamefont {Blaschke}\ and\ \citenamefont
  {Stopkowicz}(2022)}]{Blaschke2022}%
  \BibitemOpen
  \bibfield  {author} {\bibinfo {author} {\bibfnamefont {S.}~\bibnamefont
  {Blaschke}}\ and\ \bibinfo {author} {\bibfnamefont {S.}~\bibnamefont
  {Stopkowicz}},\ }\href@noop {} {\bibfield  {journal} {\bibinfo  {journal} {J.
  Chem. Phys.}\ }\textbf {\bibinfo {volume} {156}},\ \bibinfo {pages} {044115}
  (\bibinfo {year} {2022})}\BibitemShut {NoStop}%
\bibitem [{\citenamefont {London}(1937)}]{London1937}%
  \BibitemOpen
  \bibfield  {author} {\bibinfo {author} {\bibfnamefont {F.}~\bibnamefont
  {London}},\ }\href@noop {} {\bibfield  {journal} {\bibinfo  {journal} {J.
  Phys. Radium}\ }\textbf {\bibinfo {volume} {8}},\ \bibinfo {pages} {397}
  (\bibinfo {year} {1937})}\BibitemShut {NoStop}%
\bibitem [{\citenamefont {Hameka}(1958)}]{Hameka1958}%
  \BibitemOpen
  \bibfield  {author} {\bibinfo {author} {\bibfnamefont {H.~F.}\ \bibnamefont
  {Hameka}},\ }\href@noop {} {\bibfield  {journal} {\bibinfo  {journal} {Mol.
  Phys.}\ }\textbf {\bibinfo {volume} {1}},\ \bibinfo {pages} {203} (\bibinfo
  {year} {1958})}\BibitemShut {NoStop}%
\bibitem [{\citenamefont {Ditchfield}(1976)}]{Ditchfield1976}%
  \BibitemOpen
  \bibfield  {author} {\bibinfo {author} {\bibfnamefont {R.}~\bibnamefont
  {Ditchfield}},\ }\href@noop {} {\bibfield  {journal} {\bibinfo  {journal} {J.
  Chem. Phys.}\ }\textbf {\bibinfo {volume} {65}},\ \bibinfo {pages} {3123}
  (\bibinfo {year} {1976})}\BibitemShut {NoStop}%
\bibitem [{\citenamefont {Helgaker}\ and\ \citenamefont
  {J{\o}rgensen}(1991)}]{Helgaker1991}%
  \BibitemOpen
  \bibfield  {author} {\bibinfo {author} {\bibfnamefont {T.}~\bibnamefont
  {Helgaker}}\ and\ \bibinfo {author} {\bibfnamefont {P.}~\bibnamefont
  {J{\o}rgensen}},\ }\href@noop {} {\bibfield  {journal} {\bibinfo  {journal}
  {J. Chem. Phys.}\ }\textbf {\bibinfo {volume} {95}},\ \bibinfo {pages} {2595}
  (\bibinfo {year} {1991})}\BibitemShut {NoStop}%
\bibitem [{\citenamefont {Culpitt}\ \emph {et~al.}(2021)\citenamefont
  {Culpitt}, \citenamefont {Peters}, \citenamefont {Tellgren},\ and\
  \citenamefont {Helgaker}}]{Culpitt2021}%
  \BibitemOpen
  \bibfield  {author} {\bibinfo {author} {\bibfnamefont {T.}~\bibnamefont
  {Culpitt}}, \bibinfo {author} {\bibfnamefont {L.~D.~M.}\ \bibnamefont
  {Peters}}, \bibinfo {author} {\bibfnamefont {E.~I.}\ \bibnamefont
  {Tellgren}}, \ and\ \bibinfo {author} {\bibfnamefont {T.}~\bibnamefont
  {Helgaker}},\ }\href@noop {} {\bibfield  {journal} {\bibinfo  {journal} {J.
  Chem. Phys.}\ }\textbf {\bibinfo {volume} {155}},\ \bibinfo {pages} {024104}
  (\bibinfo {year} {2021})}\BibitemShut {NoStop}%
\bibitem [{\citenamefont {Peters}\ \emph {et~al.}(2022)\citenamefont {Peters},
  \citenamefont {Culpitt}, \citenamefont {Tellgren},\ and\ \citenamefont
  {Helgaker}}]{Peters2022}%
  \BibitemOpen
  \bibfield  {author} {\bibinfo {author} {\bibfnamefont {L.~D.~M.}\
  \bibnamefont {Peters}}, \bibinfo {author} {\bibfnamefont {T.}~\bibnamefont
  {Culpitt}}, \bibinfo {author} {\bibfnamefont {E.~I.}\ \bibnamefont
  {Tellgren}}, \ and\ \bibinfo {author} {\bibfnamefont {T.}~\bibnamefont
  {Helgaker}},\ }\href@noop {} {\bibfield  {journal} {\bibinfo  {journal} {J.
  Chem. Phys}\ }\textbf {\bibinfo {volume} {157}},\ \bibinfo {pages} {134108}
  (\bibinfo {year} {2022})}\BibitemShut {NoStop}%
\bibitem [{\citenamefont {Schmelcher}, \citenamefont {Cederbaum},\ and\
  \citenamefont {Meyer}(1988)}]{Schmelcher1988}%
  \BibitemOpen
  \bibfield  {author} {\bibinfo {author} {\bibfnamefont {P.}~\bibnamefont
  {Schmelcher}}, \bibinfo {author} {\bibfnamefont {L.~S.}\ \bibnamefont
  {Cederbaum}}, \ and\ \bibinfo {author} {\bibfnamefont {H.~D.}\ \bibnamefont
  {Meyer}},\ }\href@noop {} {\bibfield  {journal} {\bibinfo  {journal} {Phys.
  Rev. A}\ }\textbf {\bibinfo {volume} {38}},\ \bibinfo {pages} {6066}
  (\bibinfo {year} {1988})}\BibitemShut {NoStop}%
\bibitem [{\citenamefont {Schmelcher}\ and\ \citenamefont
  {Cederbaum}(1989)}]{Schmelcher1989}%
  \BibitemOpen
  \bibfield  {author} {\bibinfo {author} {\bibfnamefont {P.}~\bibnamefont
  {Schmelcher}}\ and\ \bibinfo {author} {\bibfnamefont {L.~S.}\ \bibnamefont
  {Cederbaum}},\ }\href@noop {} {\bibfield  {journal} {\bibinfo  {journal}
  {Phys. Rev. A}\ }\textbf {\bibinfo {volume} {40}},\ \bibinfo {pages} {3515}
  (\bibinfo {year} {1989})}\BibitemShut {NoStop}%
\bibitem [{\citenamefont {Yin}\ and\ \citenamefont {Mead}(1992)}]{Yin1992}%
  \BibitemOpen
  \bibfield  {author} {\bibinfo {author} {\bibfnamefont {L.}~\bibnamefont
  {Yin}}\ and\ \bibinfo {author} {\bibfnamefont {C.~A.}\ \bibnamefont {Mead}},\
  }\href@noop {} {\bibfield  {journal} {\bibinfo  {journal} {Theor. Chim.
  Acta}\ }\textbf {\bibinfo {volume} {82}},\ \bibinfo {pages} {397} (\bibinfo
  {year} {1992})}\BibitemShut {NoStop}%
\bibitem [{\citenamefont {Peternelj}\ and\ \citenamefont
  {Kranjc}(1993)}]{Peternelj1993}%
  \BibitemOpen
  \bibfield  {author} {\bibinfo {author} {\bibfnamefont {J.}~\bibnamefont
  {Peternelj}}\ and\ \bibinfo {author} {\bibfnamefont {T.}~\bibnamefont
  {Kranjc}},\ }\href@noop {} {\bibfield  {journal} {\bibinfo  {journal}
  {Zeitschrift f{\"{u}}r Phys. B Condens. Matter}\ }\textbf {\bibinfo {volume}
  {92}},\ \bibinfo {pages} {61} (\bibinfo {year} {1993})}\BibitemShut {NoStop}%
\bibitem [{\citenamefont {Yin}\ and\ \citenamefont {Mead}(1994)}]{Yin1994}%
  \BibitemOpen
  \bibfield  {author} {\bibinfo {author} {\bibfnamefont {L.}~\bibnamefont
  {Yin}}\ and\ \bibinfo {author} {\bibfnamefont {C.~A.}\ \bibnamefont {Mead}},\
  }\href@noop {} {\bibfield  {journal} {\bibinfo  {journal} {J. Chem. Phys.}\
  }\textbf {\bibinfo {volume} {100}},\ \bibinfo {pages} {8125} (\bibinfo {year}
  {1994})}\BibitemShut {NoStop}%
\bibitem [{\citenamefont {Schmelcher}\ and\ \citenamefont
  {Cederbaum}(1997)}]{Schmelcher1997}%
  \BibitemOpen
  \bibfield  {author} {\bibinfo {author} {\bibfnamefont {P.}~\bibnamefont
  {Schmelcher}}\ and\ \bibinfo {author} {\bibfnamefont {L.~S.}\ \bibnamefont
  {Cederbaum}},\ }\href@noop {} {\bibfield  {journal} {\bibinfo  {journal}
  {Int. J. Quantum Chem.}\ }\textbf {\bibinfo {volume} {64}},\ \bibinfo {pages}
  {501} (\bibinfo {year} {1997})}\BibitemShut {NoStop}%
\bibitem [{\citenamefont {Ceresoli}, \citenamefont {Marchetti},\ and\
  \citenamefont {Tosatti}(2007)}]{Ceresoli2007}%
  \BibitemOpen
  \bibfield  {author} {\bibinfo {author} {\bibfnamefont {D.}~\bibnamefont
  {Ceresoli}}, \bibinfo {author} {\bibfnamefont {R.}~\bibnamefont {Marchetti}},
  \ and\ \bibinfo {author} {\bibfnamefont {E.}~\bibnamefont {Tosatti}},\
  }\href@noop {} {\bibfield  {journal} {\bibinfo  {journal} {Phys. Rev. B}\
  }\textbf {\bibinfo {volume} {75}},\ \bibinfo {pages} {161101} (\bibinfo
  {year} {2007})}\BibitemShut {NoStop}%
\bibitem [{\citenamefont {Berry}(1984)}]{Berry1984}%
  \BibitemOpen
  \bibfield  {author} {\bibinfo {author} {\bibfnamefont {M.~V.}\ \bibnamefont
  {Berry}},\ }\href@noop {} {\bibfield  {journal} {\bibinfo  {journal} {Proc.
  R. Soc. Lond. A}\ }\textbf {\bibinfo {volume} {392}},\ \bibinfo {pages} {45}
  (\bibinfo {year} {1984})}\BibitemShut {NoStop}%
\bibitem [{\citenamefont {Mead}(1992)}]{Mead1992}%
  \BibitemOpen
  \bibfield  {author} {\bibinfo {author} {\bibfnamefont {C.~A.}\ \bibnamefont
  {Mead}},\ }\href@noop {} {\bibfield  {journal} {\bibinfo  {journal} {Rev.
  Mod. Phys.}\ }\textbf {\bibinfo {volume} {64}},\ \bibinfo {pages} {51}
  (\bibinfo {year} {1992})}\BibitemShut {NoStop}%
\bibitem [{\citenamefont {Anandan}, \citenamefont {Christian},\ and\
  \citenamefont {Wanelik}(1997)}]{Anandan1997}%
  \BibitemOpen
  \bibfield  {author} {\bibinfo {author} {\bibfnamefont {J.}~\bibnamefont
  {Anandan}}, \bibinfo {author} {\bibfnamefont {J.}~\bibnamefont {Christian}},
  \ and\ \bibinfo {author} {\bibfnamefont {K.}~\bibnamefont {Wanelik}},\
  }\href@noop {} {\bibfield  {journal} {\bibinfo  {journal} {Am. J. Phys.}\
  }\textbf {\bibinfo {volume} {65}},\ \bibinfo {pages} {180} (\bibinfo {year}
  {1997})}\BibitemShut {NoStop}%
\bibitem [{\citenamefont {Resta}(2000)}]{Resta2000}%
  \BibitemOpen
  \bibfield  {author} {\bibinfo {author} {\bibfnamefont {R.}~\bibnamefont
  {Resta}},\ }\href@noop {} {\bibfield  {journal} {\bibinfo  {journal} {J.
  Phys. Condens. Matter}\ }\textbf {\bibinfo {volume} {12}},\ \bibinfo {pages}
  {R107} (\bibinfo {year} {2000})}\BibitemShut {NoStop}%
\bibitem [{\citenamefont {Culpitt}\ \emph {et~al.}(2022)\citenamefont
  {Culpitt}, \citenamefont {Peters}, \citenamefont {Tellgren},\ and\
  \citenamefont {Helgaker}}]{Culpitt2022}%
  \BibitemOpen
  \bibfield  {author} {\bibinfo {author} {\bibfnamefont {T.}~\bibnamefont
  {Culpitt}}, \bibinfo {author} {\bibfnamefont {L.~D.~M.}\ \bibnamefont
  {Peters}}, \bibinfo {author} {\bibfnamefont {E.~I.}\ \bibnamefont
  {Tellgren}}, \ and\ \bibinfo {author} {\bibfnamefont {T.}~\bibnamefont
  {Helgaker}},\ }\href@noop {} {\bibfield  {journal} {\bibinfo  {journal} {J.
  Chem. Phys.}\ }\textbf {\bibinfo {volume} {156}},\ \bibinfo {pages} {044121}
  (\bibinfo {year} {2022})}\BibitemShut {NoStop}%
\bibitem [{\citenamefont {Zabalo}, \citenamefont {Dreyer},\ and\ \citenamefont
  {Stengel}(2022)}]{Zabalo2022}%
  \BibitemOpen
  \bibfield  {author} {\bibinfo {author} {\bibfnamefont {A.}~\bibnamefont
  {Zabalo}}, \bibinfo {author} {\bibfnamefont {C.~E.}\ \bibnamefont {Dreyer}},
  \ and\ \bibinfo {author} {\bibfnamefont {M.}~\bibnamefont {Stengel}},\
  }\href@noop {} {\bibfield  {journal} {\bibinfo  {journal} {Phys. Rev. B}\
  }\textbf {\bibinfo {volume} {105}},\ \bibinfo {pages} {094305} (\bibinfo
  {year} {2022})}\BibitemShut {NoStop}%
\bibitem [{\citenamefont {Smith}, \citenamefont {Palke},\ and\ \citenamefont
  {Gerig}(1992)}]{Smith1992}%
  \BibitemOpen
  \bibfield  {author} {\bibinfo {author} {\bibfnamefont {S.~A.}\ \bibnamefont
  {Smith}}, \bibinfo {author} {\bibfnamefont {W.~E.}\ \bibnamefont {Palke}}, \
  and\ \bibinfo {author} {\bibfnamefont {J.~T.}\ \bibnamefont {Gerig}},\
  }\href@noop {} {\bibfield  {journal} {\bibinfo  {journal} {{Concepts Magn.
  Reson.}}\ }\textbf {\bibinfo {volume} {4}},\ \bibinfo {pages} {107} (\bibinfo
  {year} {1992})}\BibitemShut {NoStop}%
\bibitem [{Lon()}]{London}%
  \BibitemOpen
  \href@noop {} {\enquote {\bibinfo {title} {{LONDON, a quantum-chemistry
  program for plane-wave/GTO hybrid basis sets and finite magnetic field
  calculations. By E. Tellgren (primary author), T. Helgaker, A. Soncini, K. K.
  Lange, A. M. Teale, U. Ekstr{\"o}m, S. Stopkowicz, J. H. Austad, and S. Sen.
  See londonprogram.org for more information.}}}\ }\BibitemShut {NoStop}%
\bibitem [{\citenamefont {Dunning}(1989)}]{Dunning1989}%
  \BibitemOpen
  \bibfield  {author} {\bibinfo {author} {\bibfnamefont {T.~H.}\ \bibnamefont
  {Dunning}},\ }\href@noop {} {\bibfield  {journal} {\bibinfo  {journal} {J.
  Chem. Phys.}\ }\textbf {\bibinfo {volume} {90}},\ \bibinfo {pages} {1007}
  (\bibinfo {year} {1989})}\BibitemShut {NoStop}%
\bibitem [{qua()}]{quadpy}%
  \BibitemOpen
  \href@noop {} {\enquote {\bibinfo {title} {{quadpy 0.16.14, Numerical
  integration, quadrature for various domains. By N. Schlömer See
  https://pypi.org/project/quadpy/ for more information.}}}\ }\BibitemShut
  {NoStop}%
\bibitem [{\citenamefont {Pauling}(1932)}]{Pauling1932}%
  \BibitemOpen
  \bibfield  {author} {\bibinfo {author} {\bibfnamefont {L.}~\bibnamefont
  {Pauling}},\ }\href@noop {} {\bibfield  {journal} {\bibinfo  {journal} {J.
  Am. Chem. Soc.}\ }\textbf {\bibinfo {volume} {54}},\ \bibinfo {pages} {3570}
  (\bibinfo {year} {1932})}\BibitemShut {NoStop}%
\bibitem [{\citenamefont {Richter}, \citenamefont {Duarte},\ and\ \citenamefont
  {Bruns}(2021)}]{Richter2021}%
  \BibitemOpen
  \bibfield  {author} {\bibinfo {author} {\bibfnamefont {W.~E.}\ \bibnamefont
  {Richter}}, \bibinfo {author} {\bibfnamefont {L.~J.}\ \bibnamefont {Duarte}},
  \ and\ \bibinfo {author} {\bibfnamefont {R.~E.}\ \bibnamefont {Bruns}},\
  }\href@noop {} {\bibfield  {journal} {\bibinfo  {journal} {J. Chem. Inf.
  Model.}\ }\textbf {\bibinfo {volume} {61}},\ \bibinfo {pages} {3881}
  (\bibinfo {year} {2021})}\BibitemShut {NoStop}%
\end{thebibliography}

%

\end{document}